\documentclass[man,12pt,nolmodern,floatsintext,hidelinks]{apa7}
\usepackage[nodisplayskipstretch]{setspace}
\usepackage{newtxtext,newtxmath} 
\usepackage{booktabs,multirow,array}
\usepackage{graphicx}
\usepackage{enumitem}
\usepackage[nosectionbib,numberedbib]{apacite} 
\usepackage{url}
\usepackage{amsmath,amsfonts}
\usepackage{algorithm,algpseudocode}
\usepackage{tikz}
\usepackage{afterpage,lscape,pdfpages}
\usetikzlibrary{shapes,arrows.meta,decorations.markings,positioning}
\usepackage{bm} 
\newcolumntype{L}[1]{>{\raggedright\let\newline\\\arraybackslash\hspace{0pt}}m{#1}}
\newcolumntype{C}[1]{>{\centering\let\newline\\\arraybackslash\hspace{0pt}}m{#1}}
\newcolumntype{R}[1]{>{\raggedleft\let\newline\\\arraybackslash\hspace{0pt}}m{#1}}

\DeclareMathAlphabet{\mathbf}{OT1}{cmr}{b}{n}

\definecolor{red1}{RGB}{228,55,55}
\definecolor{blue1}{RGB}{60,100,178}
\definecolor{coquelicot}{rgb}{1.0, 0.22, 0.0}
\long\def\jp#1{\bgroup\color{coquelicot}#1\egroup}
\long\def\yl#1{\bgroup\color{blue1}#1\egroup}

\def\t{^\prime}

\def\eeta{\boldsymbol{\eta}}
\def\eepsilon{\boldsymbol{\epsilon}}
\def\aalpha{\boldsymbol{\alpha}}
\def\bbeta{\boldsymbol{\beta}}

\def\llambda{\boldsymbol{\lambda}}

\def\nnu{\boldsymbol{\nu}}

\def\TTheta{\mathbf{\Theta}}

\def\one{\mathbf{1}}

\def\xx{\mathbf{x}}

\def\vv{\mathbf{v}}
\def\yy{\mathbf{y}}

\def\var{\hbox{Var}}

\def\corr{\hbox{Corr}}

\def\ex{\mathsf{E}}
\def\pr{P}

\algrenewcommand\algorithmicrequire{\textbf{input:}}
\algrenewcommand\algorithmicensure{\textbf{output:}}
\algblockdefx{Do}{EndDo}{\textbf{do}}{\textbf{end do}}

\setlist[itemize, 1]{leftmargin=*, topsep=0ex, itemsep=0pt, parsep=0ex, labelindent=0pt}
\setlist[itemize, 2]{leftmargin=*, topsep=0ex, itemsep=0pt, parsep=0ex, labelindent=0pt, label=$\circ$}
\setlist[itemize, 3]{leftmargin=*, topsep=0ex, itemsep=0pt, parsep=0ex, labelindent=0pt, label=-}
\setlist[enumerate, 1]{leftmargin=*, topsep=0ex, itemsep=0pt, parsep=0ex, labelindent=0pt, label=(\alph*)}

\begin{document}
\title{Understanding Measurement Precision from a Regression Perspective}

\shorttitle{Reliability}
\authorsnames[1,2,{3,4}]{Yang Liu, Jolynn Pek, Alberto Maydeu-Olivares}
\authorsaffiliations{%
  {Department of Human Development and Quantitative Methodology\\University of Maryland, College Park},
  {Department of Psychology\\The Ohio State University},
  {Department of Psychology\\University of South Carolina},
  {Faculty of Psychology\\University of Barcelona}%
}
\authornote{Correspondence should be made to Yang Liu at
3304R Benjamin Bldg, 3942 Campus Dr, University of Maryland, College Park, MD
20742. Email: yliu87@umd.edu. The participation of Alberto Maydeu-Olivares was supported, in part, by the Research Center for Child Well-Being (NIGMS P20GM130420), grant PID2020-119755GB-I00 funded by AEI 10.13039/501100011033, and AGAUR grant 1237SGR2017.}

\abstract{%
\indent
We adopt and expand McDonald's (2011) regression framework for measurement precision, integrating two key perspectives: (a) reliability of observed scores and (b) optimal prediction of latent scores. Reliability arises from a measurement decomposition of an observed score into its true score and measurement error. In contrast, proportional reduction in mean squared error (PRMSE) arises from a prediction decomposition of a latent score into its optimal predictor (the observed expected a posteriori [EAP] score) and prediction error. Reliability is the coefficient of determination obtained by two isomorphic regressions: regressing the observed score on its true score or on all the latent variables. Similarly, PRMSE is the coefficient of determination obtained from two isomorphic regressions: regressing the latent score on its observed EAP score or all the manifest variables. A key implication of this regression framework is that both reliability and PRMSE can be estimated using a Monte Carlo (MC) method, which is particularly useful when no analytic formula is available or when the analytic calculation is involved. We illustrate these concepts with a factor analysis model and a two parameter logistic model, in which we compute reliability coefficients for different observed scores and PRMSE for different latent scores. Additionally, we provide a numerical example demonstrating how the MC method can be used to estimate reliability and PRMSE within a two-dimensional item response tree model.
  }

    \newpage
\maketitle

\setcounter{secnumdepth}{0}
Unobservable psychological constructs (e.g., personality, intelligence, attitudes) are fundamental components of psychological theories (e.g., \citeNP{cronbach&meehl.1955}). These constructs are typically operationalized as latent variables (LVs), and LVs are indicated by manifest variables (MVs; commonly referred to as observed variables, indicator variables, or items). Examples of MVs are participants' responses to items in standardized tests and survey questionnaires. Subtest or test scores might also serve as MVs. In general, MVs are designed to reflect LVs such as cognitive ability and personality. The links between LVs and MVs are specified, both conceptually and statistically, by a \textit{latent variable measurement model}.

  Responses from MVs are usually summarized by \textit{observed scores} (e.g., summed or estimated factor scores). Observed scores are frequently employed in downstream analyses such as scoring, classification, and fitting explanatory models with observed scores serving as proxies for latent scores (e.g., see \citeNP{liu&pek.2024}). In contrast to observed scores, \textit{latent scores} refer to unobservable quantities that summarize latent constructs (i.e., operationalized by LVs). For example, suppose that data are collected on a measure of extraversion, which follows a one-factor model. An example observed score is the (unweighted) summed score, which is computed by summing up all item responses. An example latent score is the LV under the one-factor model, which quantifies the extraversion construct.  

We use the broad term \emph{measurement precision} to represent the extent to which observed quantities (i.e., MVs and observed scores) and latent quantities (i.e., LVs and latent scores) align. Measurement precision can be assessed by two distinct approaches. The first approach quantifies how well observed scores are accounted for by LVs or true scores. The second approach, in contrast, quantifies how precisely latent scores are predicted by MVs or observed scores. These two approaches map onto two classes of coefficients: (a) \emph{reliability}, which quantifies the extent of variance in observed scores due to LVs or true scores and (b) \textit{proportional reduction in mean squared error} (PRMSE), which is a measure of how latent scores can be optimally predicted by MVs or observed scores. We acknowledge that PRMSE has been referred to as reliability by certain authors (e.g., \citeNP{haberman&sinharay.2010}; \citeNP{Liu2025}); however, we explicitly differentiate reliability from PRMSE here to emphasize their conceptual distinction. In practice, it is essential to report either reliability or PRMSE, for the reason that measurement imprecision not only results in mismatches between latent and observed scores but also incurs biased inference in subsequent statistical analysis (\citeNP{bollen.1989}, Chapter 5; \citeNP{cole&preacher.2014}).

  In the current paper, we introduce a unified perspective on these two distinct classes of coefficients, provide more clarity on the interpretations of reliability versus PRMSE, and suggest a novel approach to their computation. Our conceptualization is motivated by the seminal work of \citeA{mcdonald.2011}, which we further extend. Based in regression, reliability of an observed score is the coefficient of determination obtained from regressing an observed score onto LVs or true scores. Conversely, the PRMSE of a latent score is the coefficient of determination obtained from regressing a latent score onto MVs or observed scores. Both reliability and PRMSE are normalized, meaning they are bounded by zero and one. A reliability coefficient of zero implies that the observed score carries only measurement error while a value of one indicates that the observed score perfectly reflects the LVs. A PRMSE of zero implies that MVs or the observed score cannot predict the latent score whereas a PRMSE of one indicates that the latent score can be perfectly predicted from MVs.

  Our regression-based framework of measurement precision offers two key contributions. First, the framework serves as a structured guide for organizing and interpreting various normalized indices of measurement precision. Methodological developments in quantifying measurement precision are numerous and nuanced (e.g., \citeNP{revelle&condon.2019}), making it challenging for researchers to recognize subtle differences among extant coefficients. The proposed framework establishes the link between measurement precision and regression. Second, while estimating measurement precision coefficients is relatively straightforward for unidimensional linear and nonlinear measurement models, it becomes much more complex for advanced models such as multidimensional item response theory (IRT) models. Using a regression-based perspective naturally leads to adopting a Monte Carlo (MC) method to estimate reliability and PRMSE coefficients, especially for situations in which analytic solutions are unavailable or computationally untenable. Thus, the MC procedure offers a practical and straightforward alternative for estimating measurement precision coefficients.

  The paper is structured as follows. We begin by describing underlying assumptions, establishing notation, and introducing two simple examples to illustrate the regression framework of measurement precision. The first example is a (linear) one-factor model and the second example is a (nonlinear) two-parameter logistic (2PL) IRT model. Next, we introduce the regression framework emphasizing the conceptual distinction between measurement and prediction decompositions. A measurement decomposition results in a reliability coefficient and a prediction decomposition maps onto a PRMSE coefficient. We then describe distinct interpretations of reliability versus PRMSE. Next, we introduce the MC procedure for estimating reliability and PRMSE coefficients. We demonstrate the MC procedure by applying it to our two simple examples, showing that MC estimates closely align with analytically derived values. Subsequently, we apply the MC procedure to estimate reliability and PRMSE coefficients for a complex measurement model, in which analytical calculations are intractable. Here, we use data from the depressive symptom scale within the Collaborative Psychiatric Epidemiological Surveys (\citeNP{magnus&liu.2022}). We conclude by discussing the implications of our work on reporting measurement precision coefficients and outline future directions of research.

\centerline{\textbf{Preliminaries}}

Our developments adopt a model-based approach, assuming that the measurement model is correctly specified in the population. We begin by assuming that the population parameters of the measurement model are known. Reliability and PRMSE coefficients depend solely on model parameters and hence are population parameters as well. In practice, parameters in a measurement model are estimated from a finite sample of MVs, resulting in sample estimates of reliability and PRMSE. Sample estimates are subject to model misspecification and sampling variability. Thus, the model should be evaluated for its fit to the data prior to calculating measurement precision coefficients. Additionally, sampling variability in reliability and PRMSE coefficients should be conveyed with confidence intervals. Nevertheless, because our primary focus is on the regression-based formulation of measurement precision and the MC-based calculation, evaluating the impact of model misfit and developing confidence intervals for reliability and PRMSE are left as future work.

\noindent\textbf{Notation and Assumptions}\\

  Let the vector $\yy$ represent the $m$ MVs, and the vector $\eeta$ represent $d$ LVs. Here, $m$ denotes the number of MVs (length of the measurement instrument) and $d$ denotes the number of LVs (dimensionality of the measurement model). To distinguish random variables (i.e., variables before they are observed) from their fixed empirical realizations (i.e., variables when they take on fixed values), we underline the random variables. In addition, we use bold letters to denote vectors and unbold letters for scalars. To illustrate, $x$ represents a fixed scalar variable, $\underline x$ denotes a random scalar variable, and $\xx$ is a fixed vector while $\underline\xx$ is a random vector. Objects are underlined only when their random nature is essential to the presentation (e.g., when the distribution of the variable is referred to).

  As general notation, let $x$ represent an \emph{observed score}. We consider all observed scores to be a function of the MVs $\yy$. Similarly, let $\xi$ represent a \emph{latent score}, which is a function of LVs $\eeta$.\footnote{An observed score can be more completely denoted by $x(\yy)$ and a latent score by $\xi(\eeta)$ to highlight that these scores are functions of MVs and LVs, respectively. We, however, simplify our notation by omitting the dependencies of observed and latent scores on MVs and LVs.} We use non-bolded letters $x$ and $\xi$ to indicate that the scores are unidimensional, whereas bolded letters $\yy$ and $\eeta$ indicate that MVs and LVs can be multidimensional. The population measurement model determines the joint distribution of MVs $\underline\yy$ and LVs $\underline\eeta$. Two conditional distributions can be derived from this joint distribution:

\begin{enumerate}
  \item Conditional distribution $\underline{x}|\eeta$. The observed score $x$ depends on the LVs $\eeta$ through this distribution. We underline the observed score $\underline{x}$ in this expression because it is a function of the MVs $\underline{\yy}$ that are random given $\eeta$. Reliability coefficients are derived from this conditional distribution.
  \item Conditional distribution $\underline{\xi}|\yy$. Although the latent score $\xi$ cannot be directly observed, it can be inferred from this distribution. We underline the latent score $\underline{\xi}$ here because it is a function of the LVs $\eeta$ that are random given $\yy$. PRMSE coefficients are derived from this conditional distribution.
\end{enumerate}

  \noindent\textbf{Regression in a General Context}\\
  Before presenting the regression framework for measurement precision, we review key features of regression with random regressors (see \citeNP{sampson.1974}). When regressors are random, the functional form of the regression is implied by the joint distribution of outcome and regressors. Consider a scalar outcome variable $\underline u$ which is regressed onto a single predictor $\underline v$. At each value of $v$, the conditional distribution $\underline u|v$ can be summarized by a measure of central tendency such as its expectation $\ex(\underline u|v)$. The regression function of $\underline u$ onto $\underline v$ is obtained by connecting these conditional expectations, forming a trace (i.e., curve) that represents the central tendency of $\underline u$ across different values of $v$ (see \citeNP{fox.2015}, p. 15). If we reverse the roles of the outcome and the regressor, the regression function becomes the trace of the conditional expectation of $\underline v$ given $u$. Notably, this reverse regression function $\ex(\underline v|u)$ is generally distinct from the regression function $\ex(\underline u|v)$. This general regression approach can be extended to accommodate multiple regressors, $\vv$. That is, regressing $\underline u$ onto $\vv$ involves tracing the conditional expectation $\ex(\underline u|\vv)$.

  The (population) \emph{coefficient of determination} summarizes the strength of the relations among the outcome and regressors. Specifically, the coefficient of determination quantifies the proportion of variability in $u$ that can be explained by the regressors $\vv$:
 \begin{equation}
   \varrho^2(\underline u, \underline\vv) = \frac{\var\big[\ex(\underline u|\vv)\big]}{\var(\underline u)}.
   \label{eq:cod}
 \end{equation}
 As $\vv$ predicts $u$ through the regression, the explained variance is reflected by the variance of the regression function $\ex(\underline u|\vv)$ (i.e., the numerator of Equation \ref{eq:cod}). Consider again the simple case of regressing scalar $u$ onto a single regressor $v$. We denote the coefficient of determination in the population by $\varrho^2(\underline u,\underline v)$. If we reverse the role of the outcome and regressor, we have $\varrho^2(\underline v, \underline u)$ instead. In general, $\varrho^2(\underline u, \underline v) \neq \varrho^2(\underline v, \underline u)$. If the regression of $u$ on $v$ and its reversal $v$ on $u$ are both linear, however, $\varrho^2(\underline u, \underline v)$ and $\varrho^2(\underline v, \underline u)$ are identical. In this special case, the two coefficients of determination are further equal to the squared (Pearson) correlation, $\varrho^2(\underline u, \underline v)= \varrho^2(\underline u, \underline v) = \corr^2(\underline u,\underline v)$. Later, we shall see that both reliability and PRMSE coefficients can be expressed as coefficients of determination.

\noindent\textbf{Example Models and Scores}\\

  We now describe two simple measurement models to illustrate the regression framework for measurement precision. The first measurement model is linear and the second measurement model is nonlinear. Estimating reliability and PRMSE coefficients for these models is straightforward, and we use these examples to also illustrate the accuracy of the MC approach.

  \noindent\textbf{\textit{Example 1: One-Factor Model}}\\

  Consider a common factor model with a single common factor (i.e., LV, which will be used interchangeably for this example) $\eta$. The $m \times 1$ MV vector $\yy$ is associated with the LV $\eta$ by the following equation:
    \begin{equation} \label{eq:fac}
      \yy = \nnu + \llambda\eta + \eepsilon,
    \end{equation}
    in which $\nnu$, $\llambda$, and $\eepsilon$ are $m\times 1$ vectors of MV intercepts, factor loadings, and unique factors, respectively. We assume that the unique factors $\eepsilon$ are uncorrelated with $\eta$, and we denote the covariance matrix of the unique factors by $\TTheta$ and the variance of $\eta$ by $\psi$. As a numerical example, consider a unidimensional one-factor model with $m=3$ MVs, in which
    \begin{equation} \label{eq:facparam}
      \nnu = \begin{bmatrix}
        0\\0\\0
      \end{bmatrix},\
      \llambda = \begin{bmatrix}
        .3\\.5\\.7
      \end{bmatrix},\ \psi = 1,\ \hbox{and }
      \TTheta = \begin{bmatrix}
        .91 & 0 & 0\\
        0 & .75 & 0\\
        0 & 0 & .51\\
      \end{bmatrix}.
    \end{equation}
     Equation \ref{eq:facparam} implies that the LV $\eta$ and the three MVs in $\yy = [y_{1}, y_{2}, y_{3}]\t$ are all standardized.

     \textbf{Observed Scores}. We consider two types of observed scores. The first observed score is a weighted sum of MVs $\yy$, called the regression factor score \cite{thomson.1936, thurstone.1935}, which we denote by
    \begin{equation} \label{eq:faceap}
      \tilde{\eta} = \frac{\llambda\t\TTheta^{-1}}{\llambda\t\TTheta^{-1}\llambda + \psi^{-1}}(\yy - \nnu).
    \end{equation}
  When the common and unique factors are normally distributed, the regression factor score coincides with the conditional mean of $\eta$ given $\yy$, $\ex(\underline\eta|\yy)$ \cite{thissen&thissen-roe.2022}. In the IRT literature, this conditional mean is referred to as the expected \textit{a posteriori} score of $\eta$; therefore, we use the term ``regression scores" and ``EAP scores" interchangeably for this one-factor example. Notationally, we use the tilde accent ``$\:\tilde{~}\:$'' to indicate the EAP score of a latent score (i.e., $\eta$ in Equation \ref{eq:faceap}). With parameter values in Equation \ref{eq:facparam},
    \begin{equation} \nonumber
        \tilde\eta = .14 y_1 +.28 y_2 + .57 y_3.
    \end{equation}
    The second observed score is the unweighted sum of MVs $\yy$, called the summed score:
    \begin{equation} \label{eq:sumsco}
      s = \one\t\yy,
    \end{equation}
    in which $\one$ is an $m\times 1$ vector of ones. For the three-item example,
    \begin{equation} \nonumber
        s = y_{1} + y_{2} + y_{3}.
    \end{equation}
    The observed scores of $\tilde{\eta}$ and $s$ are special cases of $x$.

    \textbf{Latent Scores}. Latent scores, generally represented by $\xi$, are functions of LVs, $\eeta$. Two classes of latent scores are of particular interest. The first class concerns just the LVs, expressed simply as $\eeta$.  The second class of latent scores are \emph{true scores} in Classical Test Theory \cite<CTT;>[Chapter 2]{lord&novick.1968}. The true score of $x$, denoted $\tau_x$, aligns with the scale of an observed score $x$ and is defined by
    \begin{equation} \label{eq:tscore}
      \tau_x = \ex(\underline x|\eeta).
    \end{equation}
    That is, the true score $\tau_x$ is the expectation of the observed score $\underline x$ given the LVs $\eeta$. Two clarifications are made here. First, a true score is defined for any observed score\footnote{We implicitly require that the conditional expectation exists for every $\eeta$}. Second, a true score is a latent score but a latent score may not necessarily be a true score \cite<e.g.,>{sijtsma&pfadt.2021}.

  Under the one-factor model, the latent score of interest is typically the LV itself, $\eta$. Another latent score that is often of interest is the true summed score defined as
    \begin{equation} \label{eq:factscore}
     \tau_s = \ex(\underline s|\eta) = \one\t(\nnu + \llambda\eta).
    \end{equation}
    This true summed score $\tau_s$ is sometimes referred in the literature as the expected summed score. Additionally, observe that the true summed score, $\tau_s$, is a linear transformation of $\eta$. Using the parameter values from Equation \ref{eq:facparam}, we have $\tau_s = 1.5\eta$. Thus, the latent scores $\eta$ and $\tau_s$ are perfectly correlated with each another.

 \noindent\textbf{\textit{Example 2: Two-Parameter Logistic Model}}\\

 For the second example, we consider dichotomous MVs which can take on values $k=0$ or 1 in the 2PL model \cite{birnbaum.1968} with a standard normal LV, $\eta \sim\mathcal{N}(0,1)$. The conditional probability for the $j$th MV, $\underline y_j$, taking the value $k$ given the LV $\eta$ is
\begin{equation}  \label{eq:2plirf}
  \pr(\underline{y}_j = k|\eta) = \frac{\exp\left[k(\alpha_j +  \beta_j\eta)\right]}{1 + \exp(\alpha_j + \beta_j\eta)},
 \end{equation}
  in which $\alpha_j$ and $\beta_j$ are the intercept and slope parameters for the $j$th MV, respectively. Equation \ref{eq:2plirf} is commonly referred to as the item response function.

  Additionally, we assume that conditional on the LV $\eta$, the dichotomous MVs are independent. This property, known as local independence (\citeNP{mcdonald.1994}; \citeNP{stout.2002}), implies that the conditional probability of observing $\yy = (y_1, \dots, y_m)\t$ given $\eta$ is
\begin{equation} \label{eq:clik}
  \pr(\yy|\eta) = \prod_{j=1}^m \pr(\underline{y}_j = y_j|\eta).
\end{equation}
As a numerical example, consider a test composed of $m = 3$ dichotomous MVs with parameters specified as
\begin{equation}  \label{eq:2plparam}
  \aalpha = \begin{bmatrix}
    1 \\ 0 \\ -2
  \end{bmatrix},\ \hbox{and }\bbeta = \begin{bmatrix}
    1\\ 1.5 \\ 2
  \end{bmatrix},
\end{equation}
  in which $\aalpha$ and $\bbeta$ contain the intercept and slope parameters, respectively. 

   \textbf{Observed Scores}. We consider two observed scores. First, the EAP score of $\eta$ based on the 2PL model is
\begin{equation}   \label{eq:2pleap}
  \tilde{\eta} = \ex( \underline\eta|\yy) = \frac{\int \eta \pr(\yy|\eta) \phi(\eta)d\eta}{\pr(\yy)},
\end{equation}
in which $\phi(\eta)$ denotes the standard normal density function of the LV $\eta$ and
\begin{equation}  \label{eq:mlik}
  \pr(\yy) = \int \pr(\yy|\eta)\phi(\eta)d\eta
\end{equation}
is the marginal probability of response pattern $\yy$. The integrals in Equations \ref{eq:2pleap} and \ref{eq:mlik} do not have closed-form expressions and are approximated by numerical quadrature in practice. With three binary MVs, there are $2^3 = 8$ response patterns. Using item parameter values specified in Equation \ref{eq:2plparam}, the marginal probabilities and EAP scores for these 8 response patterns are summarized in Table \ref{tab:2plsco}. The second observed score is the summed score $s$, which has the same expression as the summed score for the one-factor model in Equation \ref{eq:sumsco}: $s=\one\t\yy$. While the EAP score $\tilde{\eta}$ depends on a measurement model and requires knowledge of its parameters, summed scores $s$ can be obtained independently of any measurement model. In our numerical example, there are four distinct values of the summed score $s$ for the eight response patterns (see Table \ref{tab:2plsco}).

\begin{table}[!t]
  \centering
  \caption{Patterns of scores for $m=3$ binary items, marginal probabilities, expected \textit{a posteriori} (EAP) scores of the LV $\eta$, $\tilde{\eta}$, and summed scores, $s$, for all eight response patterns in the example 2PL model. The R package \texttt{mirt} with the default tuning setup was used to compute marginal probabilities and EAP scores.
  }
  \label{tab:2plsco}
  \begin{tabular}{cccc}
    \toprule
    Pattern & Probability & $\tilde{\eta}$ & $s$\\
    \midrule
 $[0, 0, 0]\t$ & .19 &$-0.96$ & 0 \\
 $[1, 0, 0]\t$ & .26 &$-0.41$ & 1 \\
 $[0, 1, 0]\t$ & .08 &$-0.16$ & 1 \\
 $[0, 0, 1]\t$ & .01 &$ 0.08$ & 1 \\
 $[1, 1, 0]\t$ & .24 &$ 0.31$ & 2 \\
 $[1, 0, 1]\t$ & .04 &$ 0.54$ & 2 \\
 $[0, 1, 1]\t$ & .02 &$ 0.76$ & 2 \\
 $[1, 1, 1]\t$ & .15 &$ 1.22$ & 3 \\
    \bottomrule
  \end{tabular}
\end{table}

\begin{figure}[!t]
  \centering
  \includegraphics[width=\textwidth]{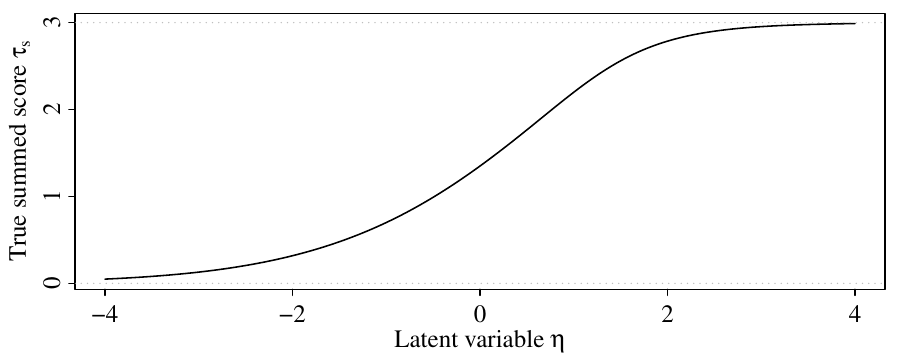}
  \caption{True summed score $\tau_s$ as a nonlinear, strictly monotone function of latent variable $\eta$ under the 2PL model (see Equation \ref{eq:2pltscore}). Item parameters used to generate the plot can be found in Equation \ref{eq:2plparam}. Horizontal dotted reference lines at 0 and 3 represent the lower and upper asymptotes of the curve.}
  \label{fig:tcc}
\end{figure}

\textbf{Latent Scores}. We consider two latent scores associated with the 2PL model. Similar to the one-factor model example, the first latent score of interest is the LV itself, $\eta$. The second latent score is the true summed score \cite{sijtsma.et.al.2024}, expressed as
    \begin{equation} \label{eq:2pltscore}
      \tau_s = \ex(\underline{s}|\eta) = \sum_{j=1}^m \pr(y_j=1|\eta),
    \end{equation}
    When plotted as a function of $\eta$, Equation \ref{eq:2pltscore} is also referred to as the test characteristic curve (\citeNP<e.g.,>[pp. 159-160]{thissen&wainer.2001}). Figure \ref{fig:tcc} presents such a graph. The true summed score $\tau_s$ is a nonlinear and strictly increasing function of $\eta$ using the slope and intercept parameters from Equation \ref{eq:2plparam}. Moreover, $\eta$ is bounded between 0 and the total number of MVs ($m=3$ in this example).

In the next sections, we review two approaches to assessing measurement precision using normalized indices, which correspond to the measurement and prediction decompositions described by \citeA{mcdonald.2011}. Both decompositions stem from regressions but differ in whether an observed score or a latent score is the outcome variable. The measurement decomposition aligns with the definition of reliability in CTT. In contrast, the prediction decomposition leads to PRMSE as a measure of measurement precision, which is more commonly employed in the item response theory (IRT) literature.

\centerline{\textbf{Measurement Decomposition}}

We begin by adapting classical definitions in Lord and Novick (\citeyearNP{lord&novick.1968}, p. 34) to the context of LV measurement models. In such models, the joint distribution of MVs $\yy$ and LVs $\eeta$ is determined by the underlying measurement model. From this joint distribution, the conditional distribution of an observed score $x$ given the LVs $\eeta$, denoted by $\underline x|\eeta$, is uniquely identified. This conditional distribution is referred to as the \emph{propensity distribution} of the observed scores $x$ (cf. \citeNP{lord&novick.1968}, p. 29--38) and captures the variability of observed scores across independent and identically distributed (i.i.d.) measurement instances of the same person. From a different perspective, the conditional distribution $\underline x|\eeta$ captures the variability of observed scores across persons with the same LV value.\footnote{See \citeA{holland.1990} for relevant discussions on the two interpretations of a propensity distribution.}

The \emph{true score} corresponding to the observed score $x$ is defined as the expectation of the propensity distribution: $\tau_x =\ex(\underline s|\eeta)$ (Equation \ref{eq:tscore}; cf. Lord \& Novick, 1968, Definition 2.3.1). Conceptually, the true score $\tau_x$ represents the long-run average of observed scores $x$ across repeated i.i.d measurements of the same individual (with a LV value $\eeta$). The true score can also be conceived as the average of observed scores within the subpopulation with specific values on the LVs $\eeta$. The true score is thus considered a version of the observed score that is free of measurement error. 

  The error score for an individual, denoted by $\varepsilon$, is the difference between the observed score and the true score:
    \begin{equation} \label{eq:escore}
      \varepsilon = x - \tau_x 
    \end{equation}
    (cf. \citeNP{lord&novick.1968}, Definition 2.4.1). This error score reflects the deviation of an individual's observed score from their true score and represents the component of variability due to measurement error. Over the population of individuals, the error score has a mean of zero and is uncorrelated with the true score. The properties of the error score are justified in Theorem 2.7.1 of \citeA{lord&novick.1968} and detailed in Section A.1 of the Supplementary Materials.

 \noindent \textbf{A Regression Formulation}\\
 
The true score $\tau_x$ and error score $\varepsilon$ in Equations \ref{eq:tscore} and \ref{eq:escore}, respectively, can be rearranged to yield the general \emph{true score formula}:
\begin{equation} \label{eq:meas}
  x = \ex(\underline x|\eeta) + \varepsilon = \tau_x + \varepsilon.
\end{equation}
 Equation \ref{eq:meas} can be expressed in words as:
    \begin{equation*}
      \hbox{observed score = true score + measurement error}.
    \end{equation*}
    \citeA{mcdonald.2011} referred to Equation \ref{eq:meas} as the \textit{measurement decomposition}, which can be viewed in two ways. First, Equation \ref{eq:meas} is a regression of the observed score $x$ onto the LVs $\eeta$: $x = \ex(\underline x|\eeta) + \varepsilon$. Second, Equation \ref{eq:meas} is a simple linear regression of the observed score $x$ onto a true score $\tau_x$: $x = \tau_x + \varepsilon$.

    In the first expression of Equation \ref{eq:meas}, the regression of observed scores $x$ onto LVs $\eeta$ can be linear or nonlinear, for which the regression function is given by $\ex(\underline x|\eeta)$ (see ``Regression in a General Context''). For instance, the separate regressions of the EAP score $\tilde\eta$ and the summed score $s$ on $\eeta$ are both linear in the one-factor example. In nonlinear measurement models or other definitions of observed scores, the regression can be nonlinear. For instance, the separate regressions of the EAP score $\tilde\eta$ and the summed score $s$ on $\eta$ are both nonlinear in the 2PL example. 

    The second expression of Equation \ref{eq:meas} treats the true score $\tau_x$ as a single regressor. In particular, this simple regression $x = \tau_x + \varepsilon$ has an intercept of zero and a slope of one. This alternative regression interpretation relies on the result that the conditional expectation of the observed score $x$ given the true score $\tau_x$ is mathematically equivalent to the conditional expectation of $x$ given the LVs $\eeta$ (i.e., $\ex(\underline x|\tau_x) = \ex(\underline x|\eeta)$), which is formally described in Section A.2 of the Supplementary Materials.

    \noindent \textbf{Definition of Reliability}\\

    Let $\mathrm{Rel}(\underline x)$ denote reliability, highlighting that it is a property of random observed scores, $\underline x$. Reliability can be defined in two equivalent ways.
 \begin{enumerate}
   \item \textbf{Ratio of variances and squared correlation}. In CTT, reliability is often be expressed as the variance of the true score $\tau_x$ divided by the variance of the observed score $x$. This variance ratio further coincides with the squared (Pearson) correlation between the observed score $x$ and its true score $\tau_x$.\footnote{Note that the second definition requires an additional assumption that $\var(\underline\tau_x) > 0$. Additionally, a third definition of reliability is the correlation between two parallel test scores. This interpretation is tangential to the current paper and thus summarized in Section B of the Supplementary Materials.} Formally, we have 
   \begin{equation}
     \mathrm{Rel}(\underline x) = \frac{\var(\underline\tau_x)}{\var(\underline x)} = \corr^2\big(\underline x, \underline \tau_x\big).
     \label{eq:reldef}
   \end{equation}
   
 \item \textbf{Coefficient of determination}. 
   Recall that the true score of $x$ traces the regression function of the observed score $x$ on the LVs $\eeta$: that is, $\tau_x = \ex(\underline x|\eeta)$. Therefore, reliability of $x$ is the coefficient of determination when regressing $x$ on $\eeta$ (Equation \ref{eq:cod}). Recall that it is the first regression expression of the  measurement decomposition (Equation \ref{eq:meas}). Moreover, because the two regression expressions of Equation \ref{eq:meas} share the same error term $\varepsilon$, reliability of $x$ further equals to the coefficient of determination when regressing $x$ on $\tau_x$. To summarize, we have
      \begin{equation}
        \mathrm{Rel}(\underline x) = \varrho^2(\underline x,\underline \eeta) = \varrho^2(\underline x, \underline\tau_x).
        \label{eq:measrel}
      \end{equation}
      In words, Equation \ref{eq:measrel} quantifies the variance in the observed score $x$ that is explained by the LVs $\eeta$, or equivalently by the true score $\tau_x$. Because the regression of $x$ on $\tau_x$ is a simple linear regression, its coefficient of determination $\varrho^2(\underline x, \underline\tau_x)$ is further equivalent to $\corr^2(\underline x, \underline\tau_x)$ (see ``Regression in a General Context'').

 \end{enumerate}

 Reliability is a property of the observed score $x$ and is based on the conceptualization that the observed score, which is a summary statistic of the MVs $\yy$, reflects the underlying LVs $\eeta$. Thus, the measurement decomposition prescribes a generative relation whereby the LVs $\eeta$ give rise to the MVs $\yy$, and the MVs are combined to form the observed score $x$. Meanwhile, the true score $\tau_x$ should be understood as the predicted value of the observed score $x$ at each LV value $\eeta$ based on the regression of $x$ onto $\eeta$. When the measurement model incorporates multiple LVs, the true score of a given observed score depends potentially on all the LVs and thus may not be a pure reflection of any single LV (cf. \citeNP{borsboom&mellenbergh.2002}).

    Next, we calculate reliability for the two examples: the one-factor model and the 2PL model. For each example, we analytically compute the reliability coefficients of the EAP score $\tilde{\eta}$ and the summed score $s$. Later, we show (in ``Estimating Reliability and PRMSE by Simulation'') how these reliability coefficients are reproduced using the MC approach, which is a direct application of the regression framework.

\noindent\textbf{\textit{Example 1: One-Factor Model}}\\

To calculate reliability analytically for an observed score $x$, we first define its true score $\tau_x$. Then, we obtain analytical formulas for $\var(\underline x)$ and $\var(\underline \tau_x)$. Finally, reliability is computed by taking the ratio of the true score variance $\var(\underline\tau_x)$ to the observed score variance $\var(\underline x)$.

  Consider the regression (EAP) factor score $\tilde{\eta}$ for the one-factor model (Equation \ref{eq:faceap}). Let $\tau_{\tilde{\eta}}$ denote the true score for this EAP score:
\begin{equation} \label{eq:facteap}
    \tau_{\tilde{\eta}} = \ex(\tilde{\underline\eta}|\eta) = \frac{\llambda\t\TTheta^{-1}\llambda}{\llambda\t\TTheta^{-1}\llambda + \psi^{-1}}\eta.
\end{equation}
This true EAP score $\tau_{\tilde{\eta}}$ is the mean of the observed EAP score $\tilde\eta$ taken across its propensity distribution. $\tau_{\tilde{\eta}}$ is unobservable and free of measurement error; it is, however, introduced for the sole purpose of computing reliability analytically. The variance of the observed EAP score $\tilde{\eta}$ is given by
    \begin{equation}   \label{eq:facvareap}
      \var(\tilde{\underline \eta}) = \frac{\psi\llambda\t\TTheta^{-1}\llambda}{\llambda\t\TTheta^{-1}\llambda + \psi^{-1}},
    \end{equation}
  and the variance of the true EAP score $\tau_{\tilde{\eta}}$ is
\begin{equation} \label{eq:facvarteap}
  \var(\underline \tau_{\tilde{\eta}}) =
  \psi\left(\frac{\llambda\t\TTheta^{-1}\llambda}{\llambda\t\TTheta^{-1}\llambda + \psi^{-1}}\right)^2.
\end{equation}
  Reliability for the EAP score $\tilde{\eta}$ is then
\begin{equation}   \label{eq:faceaprel}
  \mathrm{Rel}(\tilde{\underline \eta}) = \frac{\var(\underline \tau_{\tilde{\eta}})}{\var(\tilde{\underline \eta})} = \frac{\llambda\t\TTheta^{-1}\llambda}{\llambda\t\TTheta^{-1}\llambda + \psi^{-1}}.
\end{equation}
For our numerical one-factor example, $\mathrm{Rel}(\tilde{\underline \eta})$ is .58, implying that 58\% of the variance in the observed regression factor score $\tilde{\eta}$ is explained by the LV $\eta$ or explained by the true EAP score $\tau_{\tilde{\eta}}$. Alternatively, we can say that the squared correlation between observed and true regression factor scores is .58. It is also true that the squared correlation between the observed regression factor score and the LV is .58 because the true regression factor score is a linear transformation of the LV (Equation \ref{eq:facteap}).

Next, we compute reliability for the summed score $s$. The variance of the observed summed score $s$ and the variance of true summed score $\tau_s$ (Equation \ref{eq:factscore}) are
\begin{equation}  \label{eq:facsumvar}
    \var(\underline s) = \psi(\one\t\llambda)^2 + \mathrm{tr}(\TTheta),
\end{equation}
    and
\begin{equation} \label{eq:factsumvar}
  \var(\underline{\tau}_s) = \psi(\one\t\llambda)^2,
\end{equation}
respectively. Reliability for the summed score of the one-factor model is then
 \begin{equation} \label{eq:omega}
   \mathrm{Rel}(\underline s) =\frac{\var(\underline{\tau}_s)}{\var(\underline s)} =  \frac{\psi(\one\t\llambda)^2}{\psi(\one\t\llambda)^2 + \mathrm{tr}(\TTheta)}.
  \end{equation}
  For our numerical example, the reliability of the summed score $\mathrm{Rel}(\underline s)$ is .51: That is, 51\% of the variance in the observed summed score $s$ is explained by the LV $\eta$ or explained by the true summed score $\tau_s$. The value of .51 can also be interpreted as the squared correlation between the observed and true summed scores. As the true summed score is also a linear function of the LV (Equation \ref{eq:factscore}), we may also conclude that the squared correlation between the observed summed score and the LV is .51.

  Equation \ref{eq:omega} is widely known as coefficient omega \cite[p. 89]{mcdonald.1999}. That is, coefficient omega is the reliability of the summed score $s$ under a one-factor model. When a single LV fully explains the association among all the MVs, coefficient omega can be interpreted as the proportion of variance in the summed score $s$ explained by the LV $\eta$, or equivalently explained by the true summed score $\tau_s$. When all the factor loadings are equal, coefficient omega in Equation \ref{eq:omega} becomes coefficient alpha \cite{cronbach.1951}. Stated differently, coefficient alpha is the reliability of the summed score $s$ when the population measurement model features a single LV with equal factor loadings. Although common factor models with equal loadings are seldom applied in practice, coefficient alpha remains widely used due to two appealing properties. First, coefficient alpha can be expressed as a function of the covariance matrix of the MVs, enabling convenient computation without fitting any measurement model. Second, when the MVs are uncorrelated conditional on the LVs, coefficient alpha serves as a lower bound for the reliability of summed scores \cite[Theorem 4.4.3]{lord&novick.1968}. In other words, while coefficient alpha might not be equal to the reliability of summed scores, the exact reliability is often at least as large as coefficient alpha.

\noindent\textbf{\textit{Example 2: Two-Parameter Logistic Model}}\\
Analytical calculations for reliability tend to be less tractable for nonlinear measurement models, which we illustrate with the 2PL example. Let us first consider the observed EAP score $\tilde{\eta}$ (see Equation \ref{eq:2pleap}). We first define the true EAP score:
\begin{equation} \label{eq:2plteap}
    \tau_{\tilde{\eta}} = \ex(\tilde{\eta}|\eta)=\sum_\yy \tilde{\eta}(\yy)\pr(\yy|\eta).
\end{equation}
In Equation \ref{eq:2plteap}, we momentarily highlight the dependency of the EAP score $\tilde\eta$ on the MVs $\yy$ in the notation $\tilde\eta(\yy)$ when it appears within a sum taken across all possible response patterns $\yy$. Again, the true EAP score is introduced only for calculation purposes and is typically not of theoretical interest. Next, we derive the variances of the EAP score $\tilde{\eta}$ and the true EAP score $\tau_{\tilde{\eta}}$. The variance of the observed EAP score $\tilde{\eta}$ is
    \begin{equation} \label{eq:2plvareap}
      \var(\tilde{\underline\eta}) = \sum_{\yy}\tilde{\eta}(\yy)^2\pr(\yy) - \left[  \sum_{\yy}\tilde{\eta}(\yy)\pr(\yy) \right]^2,
    \end{equation}
    and the variance of the true EAP score $\tau_{\tilde{\eta}}$ is
    \begin{equation} \label{eq:2pltvareap}
    \var(\underline{\tau}_{\tilde{\eta}}) = \int \tau_{\tilde\eta}(\eta)^2\phi(\eta)d\eta
    - \left[ \int\tau_{\tilde\eta}(\eta)\phi(\eta)d\eta \right]^2,
    \end{equation}
    in which the notation $\tau_{\tilde\eta}(\eta)$ highlights the true EAP score as a function of the LV $\eta$. Reliability for the EAP score $\tilde{\eta}$ is then the the ratio of Equation \ref{eq:2pltvareap} over Equation \ref{eq:2plvareap}. In practice, this reliability can only be evaluated approximately because the integrals involved do not have a closed-form solution. Using numerical quadrature to approximate integrals for our example, the reliability of the EAP score $\mathrm{Rel}(\tilde{\underline{\eta}})$ is .51.\footnote{To evaluate these integrals, we use the default configuration of the R package \emph{mirt}, which employs 61 equally spaced quadrature nodes ranging from -6 to 6.} In words, 51\% of the variance in the observed EAP score $\tilde{\eta}$ is explained by the LV $\eta$ or explained by the true EAP score $\tau_{\tilde{\eta}}$. We may also say that the squared correlation between the observed and true EAP scores equals to .51. Unlike the one-factor model example, .51 is no longer the squared correlation between the observed EAP score and the LV because the true EAP score is a nonlinear function of $\eta$ under the 2PL model (Equation \ref{eq:2plteap}).

   Reliability for the summed score $s$ for the 2PL model can be calculated similarly. The variance of the summed score $s$ is respectively expressed as
    \begin{equation} \label{eq:2plsumvar}
      \var(\underline s) = \sum_{\yy}s(\yy)^2\pr(\yy) - \left[  \sum_{\yy}s(\yy)\pr(\yy) \right]^2,
    \end{equation}
    in which $s(\yy) = \one\t\yy$ (Equation \ref{eq:sumsco}) with its dependency on $\yy$ made explicit. Next, the variance of the true summed score is
  \begin{equation} \label{eq:2pltsumvar}
    \var(\underline{\tau}_s) = \int \tau_s(\eta)^2\phi(\eta)d\eta-\left[ \int \tau_s(\eta)\phi(\eta)d\eta \right]^2,
  \end{equation}
  in which we write $\tau_s(\eta)$ to indicate that the true summed score depends on the LV $\eta$.  Reliability for the summed score $s$ is the ratio of Equation \ref{eq:2pltsumvar} over Equation \ref{eq:2plsumvar}. For our numerical example, reliability of the summed score $\mathrm{Rel}(\underline s)$ is .50, implying that 50\% of the variance in the summed score $s$ is explained by the underlying LV $\eta$ or explained by the true summed score $\tau_s$. We can also say that the squared correlation between the observed and true summed scores is .50. However, it is incorrect to say that the squared correlation between the observed summed score and the LV is .50 because the true summed score $\tau_s$ is an S-shaped (i.e., nonlinear) function of $\eta$ (see Figure \ref{fig:tcc}).

\centerline{\textbf{Prediction Decomposition: Optimal Prediction of a Latent Score}}

This section focuses on quantifying measurement precision using a normalized index (bounded between 0 and 1) that measures how well a latent score $\xi$, which is a function of the LVs $\eeta$, can be predicted. Uncertainty in prediction is usually quantified by the mean squared error (MSE), and a predictor is considered optimal when it minimizes the MSE. When predicting a latent score $\xi$ using a constant $c$ that does not depend on the MVs, the MSE to be minimized is $\mathrm{MSE}(c) = \ex(\underline\xi - c)^2$. The constant that minimizes this MSE is $c=\ex\underline\xi$;\footnote{For a rigorous proof, see Casella and Berger, \citeyearNP{casella&berger.2002}, Example 2.2.6.} the MSE evaluated at this solution is $\mathrm{MSE}(c)=\var(\underline\xi)$, which represents the variance of the latent score $\xi$ across the population of individuals. This variance can be interpreted as the total amount of variability (or uncertainty) arising from individual differences in the latent score $\xi$.

Now consider predicting a latent score $\xi$ using a function of the MVs, denoted by $x(\yy)$. In the context of measurement, $x(\yy)$ is an observed score, which is dependent on the MVs $\yy$ as emphasized in the notation. We aim to find the optimal predictor (i.e., observed score) $x(\yy)$ that minimizes $\mathrm{MSE}(x) = \ex \big[\underline\xi - x(\underline\yy)\big]^2$. Such an optimal predictor among all possible functions of $\yy$ turns out to be the observed EAP score of the latent score $\xi$:
\begin{equation} \label{eq:eap}
    \tilde{\xi} = \ex(\underline\xi|\yy).\footnote{See Exercise 4.13 of \citeA{casella&berger.2002} for a justification}
\end{equation}
With the choice $x = \tilde\xi$, the minimized MSE is $\mathrm{MSE}(\tilde\xi) = \ex\big[ \var(\underline \xi|\underline\yy)\big]$; that is, the population average of the conditional variance $\var(\underline\xi|\yy)$. The value $\ex\left[ \var(\underline \xi|\underline\yy)\right]$ is the leftover uncertainty about the latent score $\xi$ after taking into account all information in the MVs $\yy$. Furthermore, the reduction in uncertainty gained by predicting the latent score $\xi$ using MVs $\yy$ (vs predicting $\xi$ with a constant $c$) is $\var(\tilde{\underline\xi}) = \var(\underline\xi) - \ex\big[ \var(\underline\xi|\underline\yy)\big]$,\footnote{This result is obtained using the law of total variance.} which reflects the degree to which the MVs $\yy$ allow for precise recovery of the latent score $\xi$. 

From this prediction perspective, measurement imprecision is couched as the uncertainty in the latent score $\xi$ that cannot be predicted using the MVs $\yy$. Let us define \emph{prediction error} as the difference between the latent score $\xi$ and its optimal predictor (i.e., the observed EAP score of $\xi$), $\delta = \xi - \tilde\xi$. When $\delta = 0$, the latent score $\xi$ can be perfectly reproduced from the MVs $\yy$ without error. In contrast, a large variance of $\delta$ means that the measurement instrument based on MVs $\yy$ is highly imprecise. Across the population of individuals, it can be shown that the prediction error $\delta$ also has a mean of zero and is uncorrelated with the observed EAP score $\tilde{\xi}$ (see Section A.1 in the Supplementary Materials). 

    \noindent \textbf{A Regression Formulation}\\

    The \emph{prediction decomposition} expresses the latent score $\xi$ as its observed EAP score $\tilde\xi$ plus the prediction error $\delta$ \cite{mcdonald.2011}:
    \begin{equation} \label{eq:pred}
      \xi = \ex(\underline\xi|\yy) + \delta = \tilde{\xi} + \delta,
    \end{equation}
    or equivalently in words,
    \begin{equation*}
      \hbox{latent score = EAP score + prediction error}.
    \end{equation*}
    The role of $\delta$ in the prediction decomposition (Equation \ref{eq:pred}) mirrors the role of $\varepsilon$ in the measurement decomposition (Equation \ref{eq:meas}). Therefore, the prediction decomposition can also be interpreted in two ways. First, Equation \ref{eq:pred} is a regression of the latent score $\xi$ onto all the MVs $\yy$: $\xi=\ex(\underline\xi|\yy) + \delta$. Second, Equation \ref{eq:pred} is a simple linear regression of the latent score $\xi$ onto its observed EAP score $\tilde{\xi}$: $\xi=\tilde{\xi} + \delta$.

  In the first expression, whether the regression of $\xi$ on $\yy$ is linear or nonlinear is fully determined by the population measurement model. In our one-factor example, $\eta$ depends on $\yy$ through a multiple linear regression and the corresponding regression function is expressed by Equation \ref{eq:faceap}. In contrast, the regression of $\eta$ on $\yy$ amounts to a linear regression with $2^m - 1$ dummy-coded patterns of MVs since there are in total $2^m$ response patterns associated with $m$ dichotomous MVs in $\yy$. This linear regression can be equivalently represented as an analysis of variance model treating each MV as a design factor and including all interaction terms (i.e., up to the $m$-way interaction).\footnote{hese conclusions hold generally for any measurement model with discrete MVs.}

    The second expression of Equation \ref{eq:pred} treats the observed EAP score $\tilde{\xi}$ as a single regressor. In particular, this simple regression $\xi = \tilde\xi + \delta$ has an intercept of zero and a slope of one. It is deduced from the result that the conditional expectation of the latent score $\xi$ given the observed EAP score $\tilde\xi$ is identical to the conditional expectation of $\xi$ given the MVs $\yy$. A justification of the latter result can be found in Section A.2 of the Supplementary Materials.

    \noindent \textbf{Definition of Proportional Reduction in Mean Squared Error}\\

    The notation $\mathrm{PRMSE}(\underline\xi)$ is used to highlight that PRMSE is a property of the random latent score $\underline\xi$. Analogous to reliability, PRMSE is subject to the following two equivalent definitions.
 \begin{enumerate}
   \item \textbf{Ratio of variances and squared correlation}. PRMSE amounts to the ratio of the variance of the observed EAP score $\tilde\xi$ to the variance of the latent score $\xi$. PRMSE is also equal to the squared (Pearson) correlation between the latent score $\xi$ and its observed EAP score $\tilde\xi$ \cite[Equation 25]{kim.2012}.\footnote{Note that the second definition requires an additional assumption that $\var(\tilde{\underline\xi}) > 0$.} Formally, 
     \begin{equation}
       \mathrm{PRMSE}(\underline\xi) = \frac{\var(\tilde{\underline\xi})}{\var(\underline\xi)} = \corr^2\big(\underline\xi, \tilde{\underline\xi}\big).
       \label{eq:prmsedef}
     \end{equation}
   \item \textbf{Coefficient of determination}. Because the observed EAP score, $\tilde\xi = \ex(\underline\xi|\yy)$, traces the regression function of the latent score $\xi$ onto the MVs $\yy$, PRMSE is simply the coefficient of determination when regressing $\xi$ onto $\yy$ (i.e., first regression expression of the prediction decomposition, Equation \ref{eq:pred}). As the same prediction error is attained by regressing $\xi$ onto its observed EAP score $\tilde\xi$ (i.e., the second regression expression of the prediction decomposition, Equation \ref{eq:pred}), we conclude that
  \begin{equation} \label{eq:predrel}
      \mathrm{PRMSE}(\underline\xi) = \varrho^2\big(\underline \xi, \underline\yy\big) =  \varrho^2\big(\underline\xi, \tilde{\underline\xi}\big).
 \end{equation}
 In words, PRMSE quantifies the proportion of variance in $\xi$ that can be explained by the MVs $\yy$, or equivalently the observed EAP score $\tilde{\xi}$, which is the optimal (i.e., MSE-minimizing) predictor of $\xi$ based on $\yy$.\footnote{Although we use the phrase ``variance can be explained'' under the prediction decomposition, we do not imply a generative relation whereby the MVs $\yy$ give rise to the latent score $\xi$.} In the meantime, the simple linear regression of $\xi$ on $\tilde\xi$ implies that $\varrho^2\big(\underline\xi, \tilde{\underline\xi}\big) = \corr^2(\underline\xi, \tilde{\underline\xi})$, which verifies the squared correlation interpretation of PRMSE.

 \end{enumerate}

  PRMSE is a property of the latent score $\xi$ and quantifies how well $\xi$ can be predicted using the MVs $\yy$. In contrast to the measurement decomposition, the prediction decomposition does not emphasize a reflective nature the measurement model. To achieve better predictive precision of a latent quantity (i.e., a latent score), one may use not only behavioral indicators (i.e., MVs) that are designed to measure $\xi$ but any additional variables that are correlated with $\xi$. We expand on the implications of measurement versus prediction decompositions in ``Measuring vs Predicting Latent Scores: Reliability vs. PRMSE''.

  In the next section, we analytically calculate PRMSE for the one-factor and 2PL examples. For each example, we calculate PMRSE for the LV itself $\eta$ and the true summed score $\tau_s$.

\noindent\textbf{\textit{Example 1: One-Factor Model}}\\
To analytically calculate PRMSE for a latent score $\xi$, we first find its observed EAP score $\tilde{\xi}$. Then, we obtain analytical formulas for $\var(\underline\xi)$ and $\var(\tilde{\underline\xi})$. Finally, PRMSE is computed by taking the ratio of the EAP score variance $\var(\tilde{\underline\xi})$ to the latent score variance $\var({\underline\xi})$.

Consider the LV $\eta$ itself as a latent score for the one-factor model. The observed EAP score $\tilde{\eta}$ is given by Equation \ref{eq:faceap}. The variance for $\eta$ is $\var(\underline\eta)=\psi$, and the variance for its observed EAP score is given in Equation \ref{eq:facvareap}. Hence, PRMSE for the LV $\eta$ is
  \begin{equation} \label{eq:faceappred}
      \mathrm{PRMSE}(\underline\eta) = \frac{\llambda\t\TTheta^{-1}\llambda}{\llambda\t\TTheta^{-1}\llambda + \psi^{-1}}.
  \end{equation}
  For our numerical one-factor example, $\mathrm{PRMSE}(\underline\eta)$ is .58, meaning that 58\% of the variance in the latent score $\eta$ is predicted by the MVs $\yy$ or predicted by the regression factor score $\tilde{\eta}$ (i.e., the EAP score of the LV). An alternative interpretation is that the squared correlation between the LV $\eta$ and regression factor score $\tilde\eta$ is .58.

  Next, we compute PRMSE for the true summed score $\tau_s$. Let the observed EAP score of the true summed score be denoted by $\tilde{\tau}_s$, which is introduced for the purpose of calculation. Recall that the true summed score $\tau_s$ is a linear transformation of LV $\eta$ (Equation \ref{eq:factscore}). Consequently, the observed EAP score of $\tau_s$ can be obtained by applying the same linear transformation to the observed EAP score of $\eta$:
    \begin{equation}
      \tilde{\tau}_s = \one\t(\nnu + \llambda\tilde\eta).
      \label{eq:eaptruesumfa}
    \end{equation}
    It follows that $\var(\underline\tau_s) = (\one\t\llambda)^2\var(\underline\eta)$ and $\var(\underline{\tilde\tau}_s) = (\one\t\llambda)^2\var(\tilde{\underline\eta})$, which implies that 
    \begin{equation}
      \mathrm{PRMSE}(\underline\tau_s) = \frac{(\one\t\llambda)^2\var(\tilde{\underline\eta})}{(\one\t\llambda)^2\var(\underline\eta)} = \mathrm{PRMSE}(\underline\eta).
      \label{eq:prmsetruesum}
    \end{equation}
  Therefore, $\mathrm{PRMSE}(\underline\tau_s)$ is also $.58$, meaning that 58\% of the variance in the true summed scores $\tau_s$ can be predicted by the MVs $\yy$ or predicted by the EAP score of the true summed score $\tilde{\tau}_s$. In addition, we conclude that the squared correlation between the true summed score $\tau_s$ and its EAP score $\tilde\tau_s$ is also .58.

  \noindent\textbf{\textit{Example 2: Two-Parameter Logistic Model}}\\

  We first consider calculating PRMSE for the LV itself $\eta$. The observed EAP score for $\eta$, $\tilde{\eta}$, was defined in Equation \ref{eq:2pleap}. The variance of the LV $\eta$ is $\var(\underline\eta)=\psi=1.0$, and the variance for its EAP score is given in Equation \ref{eq:2plvareap}. PRMSE for $\eta$ is then the ratio of these two variances. For our numerical example, $\mathrm{PRMSE}(\underline\eta)$ is .50. That is, 50\% of the variance in the latent score $\eta$ is predicted by the MVs $\yy$, or equivalently, by the EAP score $\tilde{\eta}$. It is also true that the squared correlation between the LV $\eta$ and its EAP score $\tilde\eta$ equals to .50.

  Next, consider the true summed score $\tau_s$ under the 2PL model given by Equation \ref{eq:2pltscore}. To compute the PRMSE of $\tau_s$, we need the observed EAP score of the true summed score, denoted by $\tilde\tau_s$:
\begin{equation}   \label{eq:eaptruesum2pl}
  \tilde{\tau}_s(\yy) = \ex(\underline\tau_s|\yy) = \frac{\int \tau_s(\eta) \pr(\yy|\eta) \phi(\eta)d\eta}{\pr(\yy)}
  = \frac{\sum_{j=1}^m\int \pr\big(\underline y_j = 1|\eta\big) \pr(\yy|\eta) \phi(\eta)d\eta}{\pr(\yy)},
\end{equation}
which is introduced only for computing PRMSE. The dependency of $\tilde{\tau}_s$ on $\yy$ is highlighted in the notation $\tilde{\tau}_s(\yy)$. The variance of the true summed score across the population of individuals $\var(\underline\tau_s)$ is provided in Equation \ref{eq:2pltsumvar} and has a value of .46. In addition, the variance of the EAP of the true summed score is
\begin{equation} \label{eq:2pleapvar}
  \var(\tilde{\underline{\tau}}_s) = \sum_{\yy}\tilde{\tau}_s(\yy)^2\pr(\yy)-\left[ \sum_\yy \tilde{\tau}_s(\yy)\pr(\yy) \right]^2
\end{equation}
  and has a value of .24. PRMSE for the true summed score $\tau_s$ is then computed as $\mathrm{PRMSE}(\tau_s) = $ $\var(\underline{\tilde{\tau}_s})/\var(\underline\tau_s) =$ $.24/.46$, which is .52. In words,  52\% of the variance in the true summed score $\tau_s$ can be predicted using the MVs $\yy$, or equivalently by the EAP of the true summed score $\tilde{\tau}_s$. Additionally, the squared correlation between the true summed score $\tau_s$ and its EAP score $\tilde\tau_s$ is .52.

\centerline{\textbf{Measuring vs Predicting Latent Scores: Reliability vs. PRMSE}}

Both reliability and PRMSE are normalized indices, bounded between 0 and 1, that quantify measurement precision. However, they quantify measurement precision in qualitatively distinct ways.

In a \emph{measurement decomposition}, the focus is on measuring the LVs $\eeta$ or the true score $\tau_x$ by the observed score $x$. Consequently, reliability is a property of the observed score $x$ and quantifies the extent to which observed score variance is explained by (a) the latent variables $\eeta$ or (b) the true score $\tau_x$. The true score has been defined in Equation \ref{eq:tscore} as $\tau_x =\ex(\underline x|\eeta)$, which is the regression function of the observed score $x$ on the LVs $\eeta$. Both interpretations of (a) and (b) for reliability are equivalent (see Equation \ref{eq:measrel}).

  Conversely, the \emph{prediction decomposition} perspective focuses on predicting a latent score $\xi$ by MVs $\yy$ or observed EAP scores $\tilde{\xi}$. Thus, PRMSE is a property of a latent score $\xi$ and quantifies the extent to which latent score variance can be explained by (a) the MVs $\yy$, or (b) the observed EAP score, $\tilde{\xi}$. The observed EAP score of $\xi$ is defined in Equation \ref{eq:eap} as $\tilde{\xi} = \ex(\underline \xi|\yy)$, reflecting the regression function of the latent score $\xi$ on the MVs $\yy$. Both interpretations of (a) and (b) for PRMSE are equivalent (see Equation \ref{eq:predrel}).

  In what follows, we examine the interpretations for reliability and PRMSE more closely, highlighting the roles of their underlying regressions. Additionally, we describe the special instance in which reliability and PRMSE are numerically equivalent, as well as scenarios in which one coefficient may be preferred over the other.

  \noindent\textbf{Reliability Interpretations}\\

  \noindent\textbf{\textit{Regressing an Observed Score on the Latent Variables}}\\
  Reliability reflects the proportion of variance in an observed score $x$ explained by the LVs $\eeta$. Under the population measurement model (e.g., the one-factor or 2PL model), reliability can be viewed as the coefficient of determination obtained when regressing an observed score onto all the LVs in the measurement model. The nature of this regression would depend on the model. In factor analysis, the regression would be linear whereas in the 2PL model, the regression would be nonlinear.

Table \ref{tab2:rel} summarizes the reliability coefficient values of observed scores obtained in our one-factor and 2PL examples. For the one-factor model, reliability of the regression factor score (or equivalently, the EAP score of the LV) is .58, meaning that 58\% of the variance in the these regression factor scores is attributable to individual differences in the LV. Meanwhile, the reliability of the summed score is .51, implying that 51\% of its variance is accounted for by individual differences in the LV. Comparing these two reliability coefficients, the LV explains more variability in the regression factor score compared to the summed score, which further suggests that the regression factor score is a better reflection of the LV than the summed score at the population level.

For the 2PL model, the reliability of the EAP score is .51, meaning that 51\% of the variance in the observed EAP scores can be explained by individual differences in the LV. In comparison, the reliability of the summed score is .50, implying that 50\% of the observed summed score variance is accounted for by the underlying LV. Similarly, we may conclude that the EAP score is slightly better than the summed score in reflecting the unidimensional LV under the parameter values of our population 2PL model.

An implication of our finding is that the measurement model and the scoring method must be specified for each reported reliability coefficient. Under the same measurement model (e.g., the one-factor model), there could be different reliability coefficients for different types of scores (e.g., the EAP score vs the summed scores). For the same type of score (e.g., the summed score), the reliability coefficient could be also be different when computed based on different measurement models (e.g., the one-factor model vs the 2PL model).

 \begin{table}[t!]
  \centering
  \caption{Reliability of observed scores for the one factor model and 2PL model examples: analytic and Monte Carlo (MC) estimates. Reliability estimates can be interpreted as the percentage of variance in the observed score that is explained by the latent variable (LV) or the true score. Note that EAP = expected a posteriori, and the EAP score for the one-factor model is the regression factor score.}
  \label{tab2:rel}
  \begin{tabular}{ccccc}
    \toprule
    Model & Observed outcome & Latent regressor(s) & Reliability & MC estimate\\
    \midrule
    one-factor & EAP of $\eta$ ($\tilde\eta$) & LV ($\eta$) or true score of $\tilde\eta$ ($\tau_{\tilde\eta}$) & .5821 & .5825\\
    & summed score ($s$)               & LV ($\eta$) or true summed score ($\tau_{s}$) & .5090 & .5091\\
    \midrule
    2PL & EAP of $\eta$ ($\tilde\eta$) & LV ($\eta$) or true score of $\tilde\eta$ ($\tau_{\tilde\eta}$) & .5146 & .5137\\
    & summed score ($s$) & LV ($\eta$) or true summed score ($\tau_{s}$) & .4951 & .4942\\
    \bottomrule
  \end{tabular}
\end{table}

 \noindent \textbf{\textit{Regressing an Observed Score on its True Score}}\\
 Reliability can be equivalently described as the coefficient of determination obtained when regressing an observed score $x$ onto its true score $\tau_x = \ex(\underline x|\tau_x)$. In this regression, the regression function is linear, with a slope of one and an intercept of zero (i.e., the second expression in Equation \ref{eq:measrel}). Note that a true score is a latent score because it is a function of the underlying LVs $\eeta$ in the measurement model. As such, the reliability coefficient can be alternatively described as the proportion of variance in the observed score that is explained by individual differences in the true score---the error-free counterpart of the observed score. Additionally, because this regression involves a single predictor and a linear relationship, reliability can also be defined as the squared correlation between the observed score and its true score, $\corr^2(\underline x,\underline{\tau}_x)$.

 We can therefore interpret the reliability coefficients presented in Table \ref{tab2:rel} as follows. In the one-factor example, 51\% of the variance in the summed score $s$ is explained by individual differences in the true score $\tau_s$. Equivalently, 49\% of the variance in the summed score $s$ can be attributed to measurement error. Additionally, recall for our example that the true summed score is a linear function of the LV: $\tau_s = 1.5\eta$. This one-to-one relation between $\tau_s$ and $\eta$ implies that using the true summed score versus the LV as the regressor should not change the coefficient of determination. Hence, we can also say that 51\% of the variance in the summed score $s$ is explained by the LV $\eta$. The formula for the reliability of summed scores $s$ associated with a factor analysis model is often referred to as coefficient omega (Equation \ref{eq:omega}). 

 While most readers would be familiar with true summed scores $\tau_s$, the notion of true (error-free) scores can be extended to any type of observed score $x$. Consider the observed regression factor score under the one-factor model (also known as the observed EAP score of $\eta$), $\tilde{\eta}$, presented in Equation \ref{eq:faceap}. Its error-free counterpart, the true regression factor score $\tau_{\tilde{\eta}}$ is defined in Equation \ref{eq:facteap}. From Table \ref{tab2:rel}, the reliability for the regression factor score is .58, implying that 58\% of the variance in the observed regression factor score $\tilde{\eta}$ is attributable to individual differences in the true regression factor scores $\tau_{\tilde{\eta}}$. True regression factor scores $\tau_{\tilde{\eta}}$, like true summed scores $\tau_s$, are also a linear function of the measurement model's LV $\eta$ (Equation \ref{eq:facteap}). Specifically, $\tau_{\tilde{\eta}} = .58\eta$ in the three-item numerical example. Thus, the reliability of 58\% also quantifies the amount of variance in the regression scores $\tilde{\eta}$ that is accounted for by the individual differences in the LV $\eta$.

  Reliability coefficients for the 2PL model can be interpreted in an analogous manner as those in the one-factor model as squared correlations. The observed summed score $s$ can be similarly decomposed into a true summed score $\tau_s$ and and error score. Thus, the reliability coefficient of .50 for $s$ indicates that 50\% of the variance in the observed summed scores is explained by the true summed scores $\tau_s$ (or the LV $\eta$). In the same vein, the reliability coefficient of .51 for the EAP scores $\tilde{\eta}$ indicates that 51\% of the variance in $\tilde{\eta}$ can be explained by the true EAP scores, $\tau_{\tilde{\eta}}$ (or the LV $\eta$).

The primary distinction between the one-factor and the 2PL models is whether the link between the true scores with the LV is linear versus nonlinear. For example, while the one-factor model yields a linear relation between the LV $\eta$ and the true summed score $\tau_s$, this relation in the 2PL model is nonlinear (see Figure \ref{fig:tcc}). However, regardless of the functional form relating the true score to the LVs, the relation between the observed score and its true score $\tau_x$ is always linear (see Equation \ref{eq:meas}).

\noindent\textbf{PRMSE Interpretations}\\

\noindent\textbf{\textit{Regressing a Latent Score on the Manifest Variables}}\\

PRMSE quantifies the percentage of variance in a latent score $\xi$ that can be predicted by the MVs $\yy$, from a regression of $\xi$ onto $\yy$ (see Equation \ref{eq:predrel}). Table \ref{tab3:prmse} presents PRMSE coefficients for the two numerical examples. In applications, the latent score $\xi$ of interest is usually the LV $\eta$ itself. However, functions of the LV $\eta$ might also be of interest. For example, researchers might be interested in the true summed score $\tau_s$, which is defined on the same scale as the observed summed score $s$. Thus, we focus on interpreting PRMSE of the LV itself, $\eta$, as well as PRMSE of the true summed score, $\tau_s$.

For the one-factor model, the PRMSE of the LV $\eta$ is .58, implying that 58\% of the variance in the LV $\eta$ can be explained by the continuous MVs $\yy$. Similarly, the PRMSE for the true summed score $\tau_s$, is also .58, indicating that 58\% of the variance in the true summed score $\tau_s$ can be explained by the continuous MVs $\yy$. The reason why $\mathrm{PRMSE}(\underline \eta) = \mathrm{PRMSE}(\underline\tau_s)$ for our example is because $\eta$ is perfectly correlated with $\tau_s$.

In the 2PL model, the PRMSE for the LV $\eta$ is .50, suggesting that 50\% of the variance in $\eta$ is explained by the dichotomous MVs $\yy$. Additionally, the PRMSE of the true summed score $\tau_s$ is .52, meaning that 52\% of the variance in $\tau_s$ can be accounted for by the dichotomous MVs $\yy$. The reason why $\mathrm{PRMSE}(\eta) \neq \mathrm{PRMSE}(\tau_s)$ here is because $\eta$ is nonlinearly related with $\tau_s$ for our 2PL model example as illustrated in Figure \ref{fig:tcc}.

\begin{table}[t!]
  \centering
  \caption{Proportional reduction in mean squared error (PRMSE) of latent scores for the one factor model and 2PL model examples: analytic and Monte Carlo (MC) estimates. PRMSE estimates can be interpreted as the percentage of variance in the latent score that can be explained by the manifest variables (MVs) or by the observed expected a posterori (EAP) score. Note that LV = Latent variable.}
  \label{tab3:prmse}
  \begin{tabular}{ccccc}
    \toprule
    Model & Latent outcome & Observed regressor(s) & PRMSE & MC estimate\\
    \midrule
    one-factor & LV ($\eta$) & MVs ($\yy$) or EAP of $\eta$ ($\tilde\eta$) & .5821 & .5825\\
    & true summed score ($\tau_{s}$) & MVs ($\yy$) or EAP of $\tau_s$ ($\tilde\tau_{s}$) & .5821 & .5825\\
    \midrule
    2PL & LV ($\eta$) & MVs ($\yy$) or EAP of $\eta$ ($\tilde\eta$) & .4960 & .4953\\
    & true summed score ($\tau_{s}$) & MVs ($\yy$) or EAP of $\tau_s$ ($\tilde\tau_{s}$) & .5150 & .5141\\
    \bottomrule
  \end{tabular}
\end{table}

\noindent \textbf{\textit{Regressing a Latent Score on its Observed Expected a Posteriori Score}}\\

PRMSE can be equivalently described as the coefficient of determination obtained from regressing a latent score $\xi$  onto its observed EAP score $\tilde{\xi} = \ex(\xi|\tilde{\xi})$. This regression is linear with an intercept of zero and a slope of one (i.e., the second expression in Equation \ref{eq:pred}). Note that $\tilde\xi$ is an observed score because it is a function of the MVs $\yy$ in the measurement model. Therefore, the PRMSE coefficient is alternatively interpreted as the proportion of variance in the latent score that can be explained by individual differences in its observed EAP score, which is the optimal predictor of the latent score. As the regression involves a single predictor and a linear regression function, PRMSE can also be defined as the squared correlation between the latent score and its observed EAP score, $\corr^2(\underline\xi,\underline{\tilde\xi})$.

For the one-factor model, the PRMSE for the LV $\eta$ and the true summed score $\tau_s$ are both .58, meaning that 58\% of the variance in $\eta$ and $\tau_s$ can be explained by their respective observed EAP scores $\tilde{\eta}$ and $\tilde{\tau}_s$. The two PRMSE coefficients are identical because $\eta$ is perfectly correlated with $\tau_s$, and their respective EAP scores will also be perfectly correlated.

In the same vein, the PRMSE for the LV $\eta$ is .50 under the 2PL model, meaning that 50\% of the variance in the LV can be explained by the observed EAP score $\tilde{\eta}$. In the meantime, the PRMSE for the true summed score $\tau_s$ is .52, indicating that 52\% of the variance in $\tau_s$ is explained by the observed EAP score $\tilde{\tau}_s$. The nonequivalence of the two PRMSE coefficients is due to the fact that the true summed score $\tau_s$ is a nonlinear, monotonic function of the LV $\eta$ (see Figure \ref{fig:tcc}).

It should be emphasized that PRMSE specifically requires the use of observed EAP scores $\tilde{\xi} = \ex(\underline\xi|\yy$), and not any other observed scores, as regressors for their associated latent scores $\xi$. The observed EAP scores traces the conditional expectation of the latent score $\xi$ given the MVs $\yy$ and is the optimal predictor of $\xi$ in the sense that it minimizes the predictive MSE. Thus, using observed scores as regressors other than the observed EAP score $\tilde{\xi}$ in predicting $\xi$ would result in suboptimal MSEs, reflecting an inefficient use of the information contained in the MVs $\yy$.

  \noindent\textbf{When are Reliability and PRMSE Numerically Equivalent?}

  We have noted in our one-factor numerical example that reliability of the observed EAP score of $\eta$, $\tilde\eta$, is numerically equal to PRMSE of the LV $\eta$ (i.e., .58). This is not a coincidence. In fact, we can state a necessary and sufficient condition for this equivalence: PRMSE of a latent score $\xi$ is identical to reliability of its EAP score $\tilde\xi$ if and only if the latent score $\xi$ and the true EAP score $\tau_{\tilde\xi} = \ex(\tilde\xi|\eeta)$ are perfectly correlated. A proof of the result can be found in Section C.2 of the Supplementary Materials. In the one-factor example, the true EAP score $\tau_{\tilde\eta}$ is a linear function of the LV $\eta$ (see Equation \ref{eq:facteap}), satisfying the necessary and sufficient condition. Therefore, $\mathrm{Rel}(\underline{\tilde\eta}) = \mathrm{PRMSE}(\underline\eta)$ exactly holds under the one-factor model.

  Following a similar argument, we also establish that reliability of an observed score $x$ is identical to PRMSE of its true score $\tau_x$ if and only if the observed score $x$ and the observed EAP score for the true score of $x$, $\tilde\tau_x$, are perfectly correlated. This necessary and sufficient condition explains why reliability of summed score $s$ (i.e., .51) is different from PRMSE of the true summed score $\tau_s$ (i.e., .58) in the one-factor example. Recall that the observed EAP score of the true summed score (i.e., $\tilde{\tau}_s$, Equation \ref{eq:eaptruesumfa}) is a linear transformation of the regression factor score (i.e., $\tilde\eta$, Equation \ref{eq:faceap}). Because the regression factor score does not perfectly correlate with the summed score in general, the necessary and sufficient condition is not satisfied and thus $\mathrm{PRMSE}(\underline\tau_s)\neq\mathrm{Rel}(\underline s)$.

  \noindent\textbf{Which to Report: Reliability or PRMSE?} \\

  We report both reliability and PRMSE coefficients in our examples to illustrate their computation and distinct interpretation. In practice, however, it is often unnecessary to report both reliability and PRMSE. Reliability should be reported when the research begins with an observed score $x$ and the primary interest is in assessing how well this score reflects its associated error-free (true) score $\tau_s$. In contrast, PRMSE should be reported when the research begins with an unobservable latent score $\xi$ and the focus is on evaluating how well this score can be predicted by the optimal predictor $\tilde{\xi}$. Because of the qualitative difference between reliability and PRMSE, their values cannot be sensibly compared with each other. We therefore emphasize that the choice of reporting either reliability or PRMSE (and not both) should be based on the research goal. Next, we use a hypothetical research scenario to explain this recommendation.

  Suppose that a researcher developed 10 MVs measuring impulsivity and another set of 10 MVs measuring conscientiousness. The 20 MVs in total follow a two-factor simple-structure (i.e., no cross-loading), in which the impulsivity and conscientiousness factors are negatively correlated. In one study, the researcher planned to obtain a ``pure'' observed score of impulsitivity. Because the factor model does not have cross-loadings, the researcher computed the sum of the 10 impulsivity MVs, whose true score depends only on the impulsivity LV. Here, the information provided by the 10 MVs measuring conscientiousness is ignored. In this case, reliability should be computed, and it reflects the proportion of variance in the impulsitivity summed score that can be explained solely by the impulsivity LV. In a different study, the researcher planned instead to assess how well the impulsivity LV can be predicted using observable data. In this case, the PRMSE of the impulsivity LV should be computed based on the two-factor model with all 20 MVs, including the 10 MVs measuring conscientiousness. Because conscientiousness is correlated with impulsivity, the precision of predicting the impulsitivity LV is optimized by borrowing information from the 10 MVs measuring conscientiousness.

    

\centerline{\textbf{Estimating Reliability and PRMSE by Simulation}}

Reliability and PRMSE coefficients are functions of measurement model parameters and thus can be computed analytically as illustrated with our two examples. However, analytical calculations require deriving the variances of the observed score $x$ and the true score $\tau_x$ to compute reliability, and require deriving the variances of the latent score $\xi$ and its observed EAP score $\tilde\xi$ to compute PRMSE. To simplify this process, we propose a Monte Carlo (MC) method (i.e., simulation) to estimate reliability and PRMSE coefficients. Note that both analytical and MC-based calculations of reliability and PRMSE assume known measurement model parameters and hence are subject to model misspecification and sampling variability when those parameters are estimated from sample data.

MC approaches are particularly attractive when dealing with complex nonlinear measurement models and models with discrete MVs (e.g., multidimensional IRT models). The MC method addresses two major challenges in analytical calculations. First, the MC method eliminates the need for computing integrations (numerically) with respect to LVs, which can be inefficient when the latent dimensionality is high. Second, when the MVs are discrete, the MC method avoids computing summations across all possible response patterns, which grows exponentially as the number of MVs increases and thus becomes infeasible in long tests.

Our MC procedure implements directly the regression framework by calculating reliability and PRMSE as estimated coefficients of determination (i.e., $R^2$ statistics) from fitting a regression to simulated data. For reliability, the observed score is regressed onto all the LVs. For PRMSE, the latent score is regressed onto all the MVs. The Monte Carlo procedure has three steps:
\begin{enumerate}[leftmargin=*,label={Step \arabic*}:]
  \item \textbf{Simulate LVs and MVs}. Using a measurement model with known parameters, generate a large random sample of LV vectors. For each simulated LV vector, obtain an MV vector implied by the measurement model.\\

  \item \textbf{Nonparametric regression}. To estimate reliability, compute the observed score of interest from each simulated MV vector, and then fit a nonparametric regression predicting the observed scores by the LVs. To estimate PRMSE, calculate the latent score of interest from each simulated LV vector, and then fit a nonparametric regression predicting the latent scores by the MVs. Fitting a nonparametric regression allows for linear and nonlinear functional forms between the outcome and regressor(s).\\

  \item \textbf{Estimate coefficient of determination}. Obtain the estimated coefficient of determination (i.e., the $R^2$ statistic) from the fitted regressions. From the regression of the observed scores onto the LVs, the estimated coefficient of determination would be the estimated reliability coefficient. From the regression of the latent scores onto the MVs, the coefficient of determination would be the estimated PRMSE coefficient.\\
\end{enumerate}

The regressions of observed scores (to compute reliability) and latent scores (to compute PRMSE) may be nonlinear. Nonparametric regression is handy in accommodating such nonlinearity without the need to specify exact functional forms. Additionally, the regressors in these regressions may be discrete. For instance, when estimating reliability under a latent profile analysis model, the LVs are discrete. Similarly, the MVs are discrete in IRT models when estimating PRMSE. In these cases, Step 2 reduces to fitting a linear regression with dummy-coded patterns of predictors.

In some situations, it may be computationally more convenient to replace Step 2 by:

\begin{enumerate}[leftmargin=*,label={Step 2'}:]
  \item[Step $2'$:] \textbf{Simple linear regression}. To estimate reliability, compute the observed score of interest from each simulated MV vector and its true score from the corresponding LV vector, and then fit a linear regression predicting the observed score by its true score. To estimate PRMSE, compute the latent score of interest from each simulated LV vector and its observed EAP score from the corresponding MV vector, and then fit a linear regression predicting the latent score by its observed EAP score.\\
\end{enumerate}
\noindent Steps 2 and 2$'$ are equivalent in theory. Step 2, however, is generally preferred when a nonparametric regression can be estimated with high precision (i.e., when the number of regressors are not too large). Step 2$'$ is a practical alternative when true scores for reliability and observed EAP scores for PRMSE are viable, avoiding the computational demands of high-dimensional nonparametric regression.

Reliability and PRMSE calculations are analytically straightforward in our one-factor and 2PL examples, and do not require the MC approach. However, we calculate MC-based estimates to show that they numerically match their analytical counterparts. The fifth columns of Tables \ref{tab2:rel} and \ref{tab3:prmse} present reliability and PRMSE estimates obtained from the MC procedure for the one-factor and 2PL model examples, respectively. All estimated coefficients from the MC procedure are based on a simulated sample size of $10^6$ in Step 1. From comparing the MC estimates with analytically calculated values of reliability and PRMSE coefficients, the maximum absolute difference is less than .001, indicating high accuracy of the MC-based estimates for these examples. Additionally, Step 2 (nonparametric regression) and Step 2$'$ (simple linear regression) yield estimates that are identical up to the fourth decimal in both examples.\footnote{Nonparametric regressions in Step 2 were fitted using the R package \texttt{mgcv} \cite{wood.2003}. R code for the numerical illustrations will be provided were the manuscript accepted for publication.}

\centerline{\textbf{Empirical Example}}

The simple examples described above were designed to demonstrate two key points. First, they demonstrated the distinction between the measurement and prediction decompositions of measurement precision, yielding reliability and PRMSE coefficients, respectively, and highlight the generality of the regression framework. Second, these examples also demonstrated the accuracy of the novel MC algorithm compared with analytical calculations. We now apply the MC algorithm to an empirical example to illustrate the calculation of measurement precision in a more realistic and complex measurement model, in which analytical computations are intractable.

\noindent\textbf{Data and Measurement Model}\\

  The data are from the National Comorbidity Survey Replication (NCS-R), which is part of  the Collaborative Psychiatric Epidemiological Surveys \cite{cpes}. We focus on a 14-item scale that measures depressive symptoms in the past 30 days. Each item was rated on a scale of 0 = ``never,'' 1 = ``rarely,'' 2 = ``sometimes,'' and 3 = ``often.'' In \citeA{magnus&liu.2022}, a random subset of 3000 complete responses was analyzed to illustrate the multidimensional hurdle graded response model (MH-GRM). Here, we calculate reliability and PRMSE on the same data subset (i.e., $n = 3000$ and $m = 14$), which was retrieved from \href{https://osf.io/frjm6/}{https://osf.io/frjm6/}.

  The MH-GRM was developed to account for an excessive amount of ``never'' endorsements in the data set \cite{magnus&liu.2022}, and it belongs to the family of item response tree models \cite<e.g.,>{jeon&deboeck.2016}. The MH-GRM assumes a two-stage response process for depression. The first stage represents the presence or absence of depression, and the second stage determines the severity of depression. We use two variables, $y^{(\mathrm{pres})}_j$ and $y^{(\mathrm{freq})}_j$ as recoded variables from the original 14 observed $y_j$ variables as described in Figure \ref{fig:recodedvar}. 
  \begin{figure}[!t]
  \centering
  \caption{Each ordinal item response $y_j$ = 0, 1, 2, or 3 is modeled by a two-stage decision process in the multidimensional hurdle graded response model. The first stage represents presence of a symptom, mapping onto $y_j^{(\mathrm{pres})}$ = 0 or 1. The second stage represents symptom frequency, mapping onto $y_j^{(\mathrm{freq})}$ = \texttt{NA}, 1, 2, or 3. The original $y_j$ = 0, 1, 2, and 3 can be rearranged to $(y_j^{(\mathrm{pres})}, y_j^{(2)})\t = (0, \mathtt{NA})\t$, $(1, 1)\t$, $(1, 2)\t$, and $(1, 3)\t$, respectively, in which $\mathtt{NA}$ denotes missing data.}
  \label{fig:recodedvar}
  \begin{tikzpicture}[baseline]
    \node[draw=none,inner sep=2pt](r0) at (0, 0){Never\strut};
    \node[draw=none,inner sep=2pt](r1) at (2, 0){Rarely\strut};
    \node[draw=none,inner sep=2pt](r2) at (4, 0){Sometimes\strut};
    \node[draw=none,inner sep=2pt](r3) at (6, 0){Often\strut};
    \node[draw,thick,rectangle,inner sep=3pt,minimum height=10pt]%
    (s1) at (2, 3){$y_j^{(\mathrm{pres})}$};
    \node[draw,thick,rectangle,inner sep=3pt,minimum height=10pt]%
    (s2) at (4, 1.5){$y_j^{(\mathrm{freq})}$};
    \draw[-latex,thick] (s1) -- (r0);
    \draw[-latex,thick] (s1) -- (s2);
    \node[] at (-1, 0.65) {$y_j$:\strut};
    \node[] at (0, 0.65) {0\strut};
    \node[] at (3.2, 2.4) {1\strut};
    \draw[-latex,thick] (s2) -- (r0);
    \draw[-latex,thick] (s2) -- (r1);
    \draw[-latex,thick] (s2) -- (r2);
    \draw[-latex,thick] (s2) -- (r3);
    \node[] at (1.1, 0.65) {\texttt{NA}\strut};
    \node[] at (2.5, 0.65) {1\strut};
    \node[] at (4.2, 0.65) {2\strut};
    \node[] at (5.5, 0.65) {3\strut};
  \end{tikzpicture}\\\bigskip
  \begin{tabular}{lccc}
    \toprule
    Response & Original $y_j$ & First-stage $y^{(\mathrm{pres})}_j$ & Second-stage $y^{(\mathrm{freq})}_j$\\
    \midrule
    Never     &0  &0  &\texttt{NA}\\
    Rarely    &1  &1  &1\\
    Sometimes &2  &1  &2\\
    Often     &3  &1  &3\\
    \bottomrule
  \end{tabular}
\end{figure}
A two-dimensional graded response model is applied to the recoded variables, in which indicators of symptom presence are modeled as dependent on a LV representing symptom \emph{susceptibility}, and indicators of symptom severity depend on a LV representing symptom \emph{severity}. The two LVs are assumed to be correlated. A path diagram of the MH-GRM is shown in Figure \ref{fig:mhgrm}.

In estimating the MH-GRM, \citeA{magnus&liu.2022} reported a correlation of .58 between the susceptibility and severity LVs. They also reported that observed EAP scores for susceptibility and severity exhibited different patterns in predicting health-related outcomes. For instance, when both sets of EAP scores were used in a logistic regression to predict attempted suicide, only severity scores exhibited a statistically significant partial effect. Reliability of summed and EAP scores would be informative to quantify the extent to which these observed scores reflect the susceptibility and severity LVs. Conversely, PRMSE of the susceptibility and severity LVs are useful in quantifying how accurate their EAP scores can predict their respective LVs.

\begin{figure}[!t]
  \centering
  \qquad
  \begin{tikzpicture}[baseline]
    \node[draw,thick,rectangle,inner sep=3pt,minimum height=10pt]%
    (s11) at (-1, 3){$y_{1}^{(\mathrm{pres})}$};
    \node[draw,thick,rectangle,inner sep=3pt,minimum height=10pt]%
    (s12) at (-1, 2){$y_{2}^{(\mathrm{pres})}$};
    \node[draw=none,inner sep=3pt,minimum height=10pt]%
     at (-1, 1){$\vdots$};
     \node[draw,thick,rectangle,inner sep=3pt,minimum height=10pt]%
     (s1m) at (-1, 0){$y_{4}^{(\mathrm{pres})}$};
    \node[draw,thick,rectangle,inner sep=3pt,minimum height=10pt]%
    (s21) at (9, 3){$y_{1}^{(\mathrm{freq})}$};
    \node[draw,thick,rectangle,inner sep=3pt,minimum height=10pt]%
    (s22) at (9, 2){$y_{2}^{(\mathrm{freq})}$};
    \node[draw=none,inner sep=3pt,minimum height=10pt]%
    at (9, 1){$\vdots$};
    \node[draw,thick,rectangle,inner sep=3pt,minimum height=10pt]%
    (s2m) at (9, 0){$y_{4}^{(\mathrm{freq})}$};
    \node[draw,thick,ellipse,inner sep=1pt,minimum height=8pt]%
    (l1) at (2, 1.5){\small Susceptibility};
    \node[draw,thick,ellipse,inner sep=1pt,minimum height=8pt]%
    (l2) at (6, 1.5){\small Severity};
    \draw[thick,-latex] (l1) -- (node cs:name=s11,anchor=east);
    \draw[thick,-latex] (l1) -- (node cs:name=s12,anchor=east);
    \draw[thick,-latex] (l1) -- (node cs:name=s1m,anchor=east);
    \draw[thick,-latex] (l2) -- (node cs:name=s21,anchor=west);
    \draw[thick,-latex] (l2) -- (node cs:name=s22,anchor=west);
    \draw[thick,-latex] (l2) -- (node cs:name=s2m,anchor=west);
    \draw[thick,latex-latex] (l1) ..controls (2.5, 2.7) and (5.5, 2.7).. (l2);
  \end{tikzpicture}
  \caption{Path diagram of the multidimensional hurdle graded response model. The susceptibility latent variable (LV) is indicated by the symptom presence indicators $y_{1}^{(\mathrm{pres})}, \dots, y_{14}^{(\mathrm{pres})}$, and the severity LV is indicated by the symptom frequency indicators $y_{2}^{(\mathrm{freq})}, \dots, y_{14}^{(\mathrm{freq})}$. These two LVs are allowed to covary.}
  \label{fig:mhgrm}
\end{figure}

In what follows, we use the MC procedure to calculate various reliability and PRMSE coefficients based on the fitted MH-GRM. The model parameters were estimated from a large sample of size $n = 3000$.\footnote{To apply the MC procedure, we treated the estimated model parameters as population values, ignoring the sampling variability.} Using the R package \texttt{mirt} \cite{chalmers.2012}, we fit the MH-GRM illustrated in Figure \ref{fig:mhgrm} to the recoded data represented in Figure \ref{fig:recodedvar}.\footnote{Parameter estimates were obtained using the Expectation-Maximization algorithm \cite{bock&aitkin.1981} along with default configurations of numerical quadrature and convergence criteria.}

\noindent\textbf{Reliability}

 \begin{figure}[!t]
  \centering
  \includegraphics[width=0.5\textwidth]{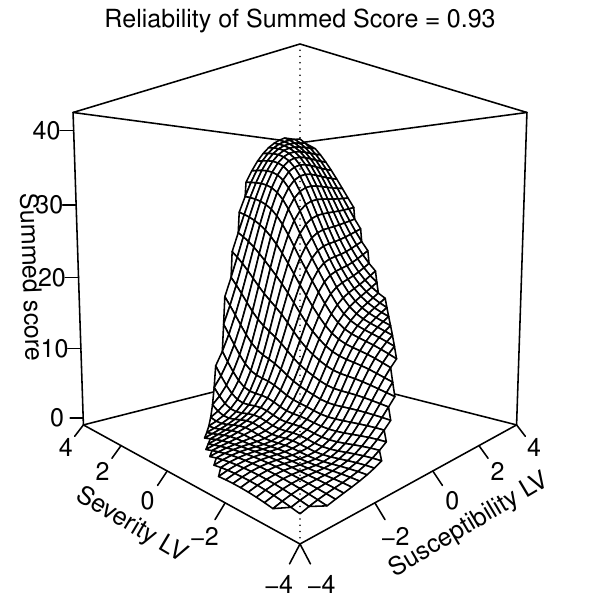}
  \caption{Regression of summed score onto susceptibility and severity latent variables. The $x$-, $z$-, and $y$-axes represent the Susceptibility LV, the Severity LV, and the summed score, respectively. Reliability of the summed score is the coefficient of determination, $R^2$.}
  \label{fig:measSS}
\end{figure}

After estimating the MH-GRM, various observed scores can be computed to reflect the susceptibility and severity LVs. Reliability coefficients provide essential information on how well observed scores reflect their underlying true scores, which are combinations of the two LVs. We consider three distinct observed scores: (a) the summed score of the original 14 items, (b) the observed EAP score of susceptibility, and (c) the observed EAP score of severity. In practice, researchers may not need to compute all three scores and or may be interested in other observed scores. To compute reliability with the MC algorithm, we simulated $10^6$ LV vectors (i.e., susceptibility and severity) from a standard bivariate normal distribution with a correlation of .58 estimated from the data. With the estimated MH-GRM parameters, we also simulated $10^6$ MV vectors (i.e., item responses to the 14 items), each of which corresponds to a simulated LV vector. From these simulated responses, we computed the summed score across the 14 MVs, and the observed EAP scores for the susceptibility and severity LVs. We then regressed each observed score onto the two LVs of susceptibility and severity. The mesh surface in Figure \ref{fig:measSS} visualizes the fitted nonparametric regression function of the summed scores ($z$-axis) onto the LVs of susceptibility ($x$-axis) and severity ($y$-axis). Values on the mesh surface represent the true summed scores across values of the susceptibility and severity LVs. 

Reliability of the summed score is estimated by the $R^2$ statistic from this regression, $R^2 = .93$, meaning that 93\% of the variance in the summed score is explained jointly by the susceptibility and severity LVs. Equivalently, this reliability coefficient indicates that 93\% of the variance in the summed score is explained by the true summed score. Alternatively, 7\% of the variance in the summed score is due to random error. This high reliability is due to the strong partial relations between both LVs and the summed score, which can be seen from Figure \ref{fig:measSS}. Note that curves on the mesh surface parallel to the axis of the susceptibility LV represents partial regressions onto the susceptibility LV given different levels of the severity LV. Similarly, partial regressions onto the severity LV are depicted by curves parallel to the axis of the severity LV. Because we see steep curves in both directions on the mesh surface in Figure \ref{fig:measSS}, we conclude that the summed score is positively and nonlinearly related to both LVs
  
  \begin{figure}[!t]
  \centering
  \includegraphics[width=\textwidth]{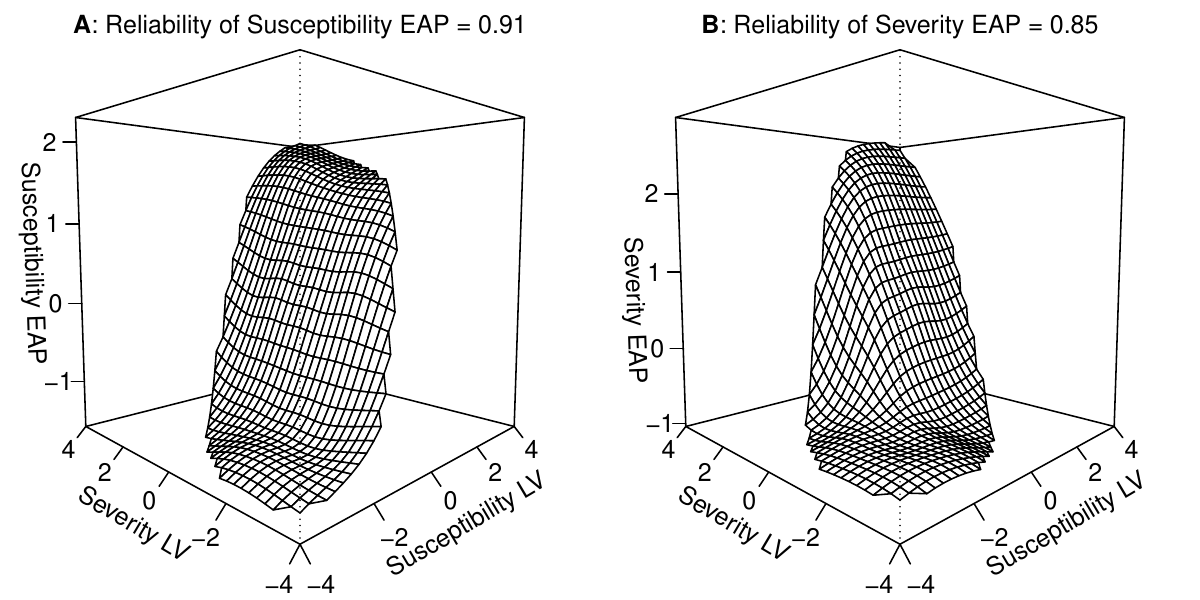}
  \caption{Regression of expected a posteriori (EAP) scores onto susceptibility and severity latent variables (LVs). Panel A: regression of EAP susceptibility score. Panel B: regression of EAP severity score. The $x$-, $z$-, and $y$-axes represent the susceptibility LV, the severity LV, and the EAP scores, respectively. Reliability of the EAP scores are the coefficient of determination, $R^2$.}
  \label{fig:measEAP}
\end{figure}

Reliability of the observed EAP score of susceptibility is .91 and reliability of the observed EAP score of severity is .85. The first reliability coefficient means that 91\% of the variance in the observed susceptibility EAP score is explained jointly by the susceptibility and severity LVs. Equivalently, 91\% of the variance in this observed EAP score is explained by its underlying true EAP score. The second reliability coefficient means that 85\% of the variance in the observed severity EAP score is explained by both the susceptibility and severity LVs, or equivalently by its underlying true EAP score. Mesh-surface plots visualizing the corresponding regression functions are shown in Figure \ref{fig:measEAP}. The regression of the observed susceptibility EAP score is presented in Figure \ref{fig:measEAP}A and the regression of the observed susceptibility EAP score is presented in \ref{fig:measEAP}B. Curves on the mesh surfaces provide information on why these reliability coefficients are different. It can be inferred from Figure \ref{fig:measEAP}A that the high reliability of the observed susceptibility EAP score (i.e., .91) is mainly attributed to the strong, positive, and nonlinear partial relation between this observed EAP score and the susceptibility LV. This is illustrated by the steeply rising curves parallel to the axis of the susceptibility LV in Figure \ref{fig:measEAP}A. Meanwhile, a weaker partial relation between the observed susceptibility EAP score and the severity LV is portrayed by nearly flat curves parallel to the axis of the severity LV. Figure \ref{fig:measEAP}B illustrates the regression function underlying the reliability coefficient .85 for the observed severity EAP score. It can be observed by tracing curves on the mesh surface that the observed severity EAP score is partially (nonlinearly) related to both LVs. However, the curves tend to be flat in both directions when both LVs are low, which likely causes a lower overall coefficient of determination compared to that of the observed susceptibility EAP score.


  \noindent\textbf{PRMSE}

\begin{figure}[!t]
  \centering
  \includegraphics[width=0.95\textwidth]{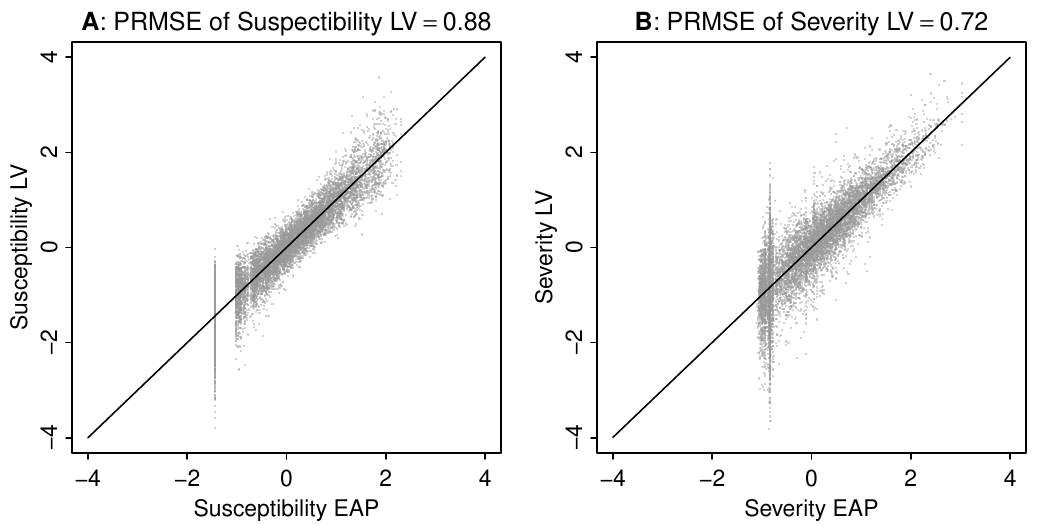}
  \caption{Regression of susceptibility and severity latent variables onto their expected a posteriori (EAP) scores. Panel A: regression of susceptibility LV. Panel B: regression of severity LV . A random subset of $10^4$ Monte Carlo (MC) samples are selected and the corresponding scores are plotted (in gray dots). Predicted values of the simple linear regressions are depicted as solid lines. Estimated PRMSE for each latent score is displayed on top of each plot as the squared correlation (coefficient of determination), $R^2$.}
  \label{fig:pred}
\end{figure}

  When interest is in how well latent scores can be predicted, focus should be placed on the PRMSE. PRMSE quantifies the amount of variability in the latent score that can be accounted for by the MVs, or equivalently by the optimal predictor of the latent score---its observed EAP score. In this application, the two primary latent scores of interest are the susceptibility and severity LVs. To compute PRMSE with the MC algorithm, we first simulated a large MC sample of $10^6$ LV and MV vectors similar to what we did for computing reliability. Then, we could either regress each of the LVs onto all the MVs or onto the respective observed EAP scores (see Equation \ref{eq:pred}). The regression of an LV onto the 14 MVs would involve $4^{14}$ regressors to indicate the $4^{14}$ distinct response patterns. Readers familiar with ANOVA can view each MV as a factor with four levels, and the regression of the latent score onto the MVs involves fitting a full 14-way factorial ANOVA including all interactions.  To avoid such a large number of regressors, we regressed the LVs onto their respective EAP scores to calculate PRMSE instead. The scatterplots in Figure \ref{fig:pred} provide a visualization of the simple linear regression of the susceptibility and severity LVs ($y$-axis) regressed onto their respective EAP scores ($x$-axis).\footnote{EAP scores were computed using the \texttt{mirt} package with the default quadrature setting. Points in Figure \ref{fig:pred} are a random subset of $10^4$ samples from the total $10^6$ MC draws.}

PRMSE of the susceptibility and the severity LVs are .88 and .72, respectively. These values mean that 88\% of the variance in the susceptibility LV can be explained by the observed EAP score of susceptibility, or equivalently by the 14 MVs. Similarly, 72\% of the variance in the severity LV can be explained by the observed EAP score of severity, or equivalently by the 14 MVs. Taken together, we conclude that the construct of susceptibility (symptom presence) can be better  predicted from the MH-GRM model than the construct of severity (symptom frequency).

The scatterplots and regression lines between the two LVs and the corresponding observed EAP scores are presented in Figures \ref{fig:pred}A and \ref{fig:pred}B, providing graphical representations of the simple regression functions underlying the two PRMSE coefficients. Observe that the regressions of latent scores onto the associated observed EAP scores are linear with an intercept of zero and a slope of one (as indicated in Equation \ref{eq:pred}). The difference in PRMSE for the susceptibility and severity LVs (.88 vs .72) is reflected by a generally larger variability of severity LVs across the entire range of their respective EAP scores. Additionally, the variability of the severity LV is high when the observed severity EAP score falls within the vicinity of $-1$.

In sum, we illustrated the how reliability and PRMSE can be calculated using the MC procedure with a complex MH-GRM measurement model for NCS-R depressive symptom data. Reliability coefficients were estimated by the MC sample coefficient of determination obtained from regressing observed scores onto the LVs. PRMSE coefficients were estimated by the MC sample coefficient of determination obtained from regressing latent scores onto their observed EAP scores. We also emphasize the different interpretations of reliability and PRMSE, highlighting that reliability is a property of an observed score whereas PRMSE is a property of a latent score. Furthermore, we show that visualizing regressions underlying reliability and PRMSE coefficients can facilitate interpreting the relations among their respective observed and latent quantities.

\centerline{\textbf{Discussion}}

In the current paper, we propose a broader understanding of coefficients of measurement precision (i.e., reliability and PRMSE) from a regression perspective. We now elaborate on a number of important implications of our work.

Reliability is not an inherent property of a measurement instrument but a property of observed scores under the true population measurement model that links LVs to MVs. Consequently, for a given set of MVs (e.g., a test or survey), the reliability of an observed score (summarizing information in the MVs) depends on two key factors: (a) the true measurement model and (b) the choice of the observed score. Consider for example a test composed of items (MVs) scored on a five-point Likert-type scale, which follow an ordinal one-factor model in the population. With a particular chosen observed score, say, the summed score, its population reliability coefficient amounts to the coefficient of determination when regressing the summed score onto all the LVs in the ordinal factor model. However, some researchers might regard the Likert-type response scale as an approximate interval scale and fit a linear factor analysis model \cite{rhemtulla.el.al.2012}. How well the linear factor model approximates the ordinal factor model would influence how close the reliability coefficients for summed scores based on these two models are. Additionally, under the ordinal factor analysis model, different scores derived from the same set of MVs (e.g., summed scores and factor scores) would generally exhibit different reliability coefficients.
  
Likewise, PRMSE is not an inherent property of a measurement instrument, but a property of latent scores under the true population measurement model. PRMSE depends on two factors as well: (a) the true measurement model and (b) the choice of the latent score. Consider again the Likert-type response scale, in which the population measurement model is the ordinal one-factor model. Under the population model, different latent scores, such as the LV itself versus the true summed score, will generally have different PRMSE coefficients. If we decide to fit a linear factor model as an approximation to the ordinal factor model, then we would conclude, as we did in our linear one-factor example, that the LV and the true summed score share the same PRMSE. This conclusion, however, could be misleading when the incorrectly specified measurement model (i.e., the linear one-factor model) is not a close approximation to the true measurement model (i.e., the ordinal one-factor model).
    
In this article, we have adopted and expanded the regression framework for measurement precision as proposed by \citeA{mcdonald.2011}. Our framework clarifies the definitions of reliability and PRMSE. Reliability of an observed score is: (a) the coefficient of determination obtained from regressing the observed score on its true score, and (b) the coefficient of determination obtained from regressing the observed score on all the LVs in the measurement model. Importantly, these two coefficients of determination are equivalent. A corollary to this dual definition is that reliability of an observed score is: (a) the proportion of variance in the observed score that is explained by its true score variance, (b) the proportion of variance in the observed score that is explained by all the model's LVs, and (c) the squared correlation between the observed score and its true score.
  
Analogously, PRMSE of a latent score is: (a) the coefficient of determination obtained from regressing the latent score on its observed EAP score, and (b) the coefficient of determination obtained from regressing the latent score on all the MVs in the measurement model. These two coefficients of determination are also equivalent. It follows from this dual definition that PRMSE of a latent score is (a) the proportion of variance in the latent score that can be explained by its EAP score, (b) the proportion of variance in the latent score that can be explained by all the model's MVs, and (c) the squared correlation between the latent score and its observed EAP score.
  
An immediate consequence of the above definitions is that the reliability of any observed score and the PRMSE of any latent score can be estimated using MC. The proposed MC procedure simulates LVs and MVs based on known measurement model parameters and calculates reliability and PRMSE as coefficients of determination from their respective regressions. We have shown that the MC procedure can yield similar results to analytic calculations in the two simple examples wherein analytical formulas are available. More importantly, the MC procedure has the advantage of being able to estimate the reliability of an observed score and PRMSE for a latent score for which no analytic formula has yet been derived or when the analytic estimate is computationally onerous. Although simulation methods have been used to assess measurement precision in complex IRT models \cite<e.g.,>{brown&maydeu.2011}, the regression perspective we formulate in the present paper provides the underlying theory to justify MC-based calculations of reliability and PRMSE.
   
The regression framework also highlights the central role of true scores in defining reliability, and analogously the central role of the observed EAP scores in defining PRMSE. Reliability arises from the measurement decomposition of an observed score into its true score and measurement error (Equation \ref{eq:meas}) whereas PRMSE arises from the prediction decomposition of a latent score into its observed EAP score and prediction error (Equation \ref{eq:pred}). The true score in a measurement decomposition summarizes the information in the LVs reflected by the MVs. Meanwhile, the observed EAP scores in a prediction decomposition amounts to the optimal (i.e., MSE-minimizing) predictor of the latent score of interest.
  
Several related topics remain for future development. First, the performance of the MC procedure should be broadly examined under a wider range of simulation conditions, including different measurement models, sample sizes, and model error, as well as across a broader set of empirical examples. We anticipate that the optimal tuning of the MC algorithm (e.g., the number of MC samples and the choice of regression estimators) will vary case by case. Second, in our empirical example, we assumed that the measurement model is correctly specified and treated the estimated parameters as their population values. In practice, however, there might be model approximation error to various extents, and small sample calibrations of measurement models may result in nonignorable sampling variability in estimated parameters. To address this, it would be important to investigate the impact of model misspecification and develop confidence intervals for MC-based reliability and PRMSE coefficients. Finally, to enhance our understanding of various measurement precision coefficients, it would be beneficial to establish benchmarks or guidelines for interpreting these distinct measures of reliability and PRMSE. Measurement precision is a complex topic, and we hope that the regression framework and MC-based computational procedure presented here contribute, respectively, to a more systematic understanding and interpretation of existing measurement precision indices in psychological research while also facilitating their calculation.
\bibliography{ref}

\begin{thebibliography}{}

\bibitem [\protect \citeauthoryear {%
Alegria%
, Jackson%
, Kessler%
\BCBL {}\ \BBA {} Takeuchi%
}{%
Alegria%
\ \protect \BOthers {.}}{%
{\protect \APACyear {2001-2003}}%
}]{%
cpes}
\APACinsertmetastar {%
cpes}%
\begin{APACrefauthors}%
Alegria, M.%
, Jackson, J\BPBI S.%
, Kessler, R\BPBI C.%
\BCBL {}\ \BBA {} Takeuchi, D.%
\end{APACrefauthors}%
\unskip\
\newblock
\APACrefYearMonthDay{2001-2003}{}{}.
\newblock
\APACrefbtitle {Collaborative Psychiatric Epidemiology Surveys ({CPES}),
  2001-2003 {[United States] (ICPSR 20240)}.} {Collaborative psychiatric
  epidemiology surveys ({CPES}), 2001-2003 {[United States] (ICPSR 20240)}.}
\newblock
\APAChowpublished {Inter-university Consortium for Political and Social
  Research [distributor], 2016-03-23. https://doi.org/10.3886/ICPSR20240.v8}.
\PrintBackRefs{\CurrentBib}

\bibitem [\protect \citeauthoryear {%
Birnbaum%
}{%
Birnbaum%
}{%
{\protect \APACyear {1968}}%
}]{%
birnbaum.1968}
\APACinsertmetastar {%
birnbaum.1968}%
\begin{APACrefauthors}%
Birnbaum, A.%
\end{APACrefauthors}%
\unskip\
\newblock
\APACrefYearMonthDay{1968}{}{}.
\newblock
{\BBOQ}\APACrefatitle {Some latent train models and their use in inferring an
  examinee's ability} {Some latent train models and their use in inferring an
  examinee's ability}.{\BBCQ}
\newblock
\BIn{} F\BPBI M.~Lord\ \BBA {} M\BPBI R.~Novick\ (\BEDS), \APACrefbtitle
  {Statistical theories of mental test scores} {Statistical theories of mental
  test scores}\ (\BPGS\ 395--479).
\newblock
\APACaddressPublisher{Reading, MA}{Addison-Wesley}.
\PrintBackRefs{\CurrentBib}

\bibitem [\protect \citeauthoryear {%
Bock%
\ \BBA {} Aitkin%
}{%
Bock%
\ \BBA {} Aitkin%
}{%
{\protect \APACyear {1981}}%
}]{%
bock&aitkin.1981}
\APACinsertmetastar {%
bock&aitkin.1981}%
\begin{APACrefauthors}%
Bock, R\BPBI D.%
\BCBT {}\ \BBA {} Aitkin, M.%
\end{APACrefauthors}%
\unskip\
\newblock
\APACrefYearMonthDay{1981}{}{}.
\newblock
{\BBOQ}\APACrefatitle {Marginal maximum likelihood estimation of item
  parameters: Application of an {EM} algorithm} {Marginal maximum likelihood
  estimation of item parameters: Application of an {EM} algorithm}.{\BBCQ}
\newblock
\APACjournalVolNumPages{Psychometrika}{46}{4}{443--459}.
\newblock
\begin{APACrefDOI} \doi{10.1007/bf02293801} \end{APACrefDOI}
\PrintBackRefs{\CurrentBib}

\bibitem [\protect \citeauthoryear {%
Bollen%
}{%
Bollen%
}{%
{\protect \APACyear {1989}}%
}]{%
bollen.1989}
\APACinsertmetastar {%
bollen.1989}%
\begin{APACrefauthors}%
Bollen, K\BPBI A.%
\end{APACrefauthors}%
\unskip\
\newblock
\APACrefYear{1989}.
\newblock
\APACrefbtitle {Structural equations with latent variables} {Structural
  equations with latent variables}.
\newblock
\APACaddressPublisher{}{John Wiley \& Sons}.
\PrintBackRefs{\CurrentBib}

\bibitem [\protect \citeauthoryear {%
Borsboom%
\ \BBA {} Mellenbergh%
}{%
Borsboom%
\ \BBA {} Mellenbergh%
}{%
{\protect \APACyear {2002}}%
}]{%
borsboom&mellenbergh.2002}
\APACinsertmetastar {%
borsboom&mellenbergh.2002}%
\begin{APACrefauthors}%
Borsboom, D.%
\BCBT {}\ \BBA {} Mellenbergh, G\BPBI J.%
\end{APACrefauthors}%
\unskip\
\newblock
\APACrefYearMonthDay{2002}{}{}.
\newblock
{\BBOQ}\APACrefatitle {True scores, latent variables, and constructs: A comment
  on {S}chmidt and {H}unter} {True scores, latent variables, and constructs: A
  comment on {S}chmidt and {H}unter}.{\BBCQ}
\newblock
\APACjournalVolNumPages{Intelligence}{30}{6}{505--514}.
\newblock
\begin{APACrefDOI} \doi{10.1016/S0160-2896(02)00082-X} \end{APACrefDOI}
\PrintBackRefs{\CurrentBib}

\bibitem [\protect \citeauthoryear {%
Brown%
\ \BBA {} Maydeu-Olivares%
}{%
Brown%
\ \BBA {} Maydeu-Olivares%
}{%
{\protect \APACyear {2011}}%
}]{%
brown&maydeu.2011}
\APACinsertmetastar {%
brown&maydeu.2011}%
\begin{APACrefauthors}%
Brown, A.%
\BCBT {}\ \BBA {} Maydeu-Olivares, A.%
\end{APACrefauthors}%
\unskip\
\newblock
\APACrefYearMonthDay{2011}{}{}.
\newblock
{\BBOQ}\APACrefatitle {Item response modeling of forced-choice questionnaires}
  {Item response modeling of forced-choice questionnaires}.{\BBCQ}
\newblock
\APACjournalVolNumPages{Educational and Psychological
  Measurement}{71}{3}{460--502}.
\PrintBackRefs{\CurrentBib}

\bibitem [\protect \citeauthoryear {%
Casella%
\ \BBA {} Berger%
}{%
Casella%
\ \BBA {} Berger%
}{%
{\protect \APACyear {2002}}%
}]{%
casella&berger.2002}
\APACinsertmetastar {%
casella&berger.2002}%
\begin{APACrefauthors}%
Casella, G.%
\BCBT {}\ \BBA {} Berger, R\BPBI L.%
\end{APACrefauthors}%
\unskip\
\newblock
\APACrefYear{2002}.
\newblock
\APACrefbtitle {Statistical inference} {Statistical inference}\
  (\PrintOrdinal{2nd}\ \BEd).
\newblock
\APACaddressPublisher{Pacific Grove, CA}{Duxbury}.
\PrintBackRefs{\CurrentBib}

\bibitem [\protect \citeauthoryear {%
Chalmers%
}{%
Chalmers%
}{%
{\protect \APACyear {2012}}%
}]{%
chalmers.2012}
\APACinsertmetastar {%
chalmers.2012}%
\begin{APACrefauthors}%
Chalmers, R\BPBI P.%
\end{APACrefauthors}%
\unskip\
\newblock
\APACrefYearMonthDay{2012}{}{}.
\newblock
{\BBOQ}\APACrefatitle {mirt: A multidimensional item response theory package
  for the {R} environment} {mirt: A multidimensional item response theory
  package for the {R} environment}.{\BBCQ}
\newblock
\APACjournalVolNumPages{Journal of Statistical Software}{48}{6}{1--29}.
\newblock
\begin{APACrefDOI} \doi{10.18637/jss.v048.i06} \end{APACrefDOI}
\PrintBackRefs{\CurrentBib}

\bibitem [\protect \citeauthoryear {%
Cole%
\ \BBA {} Preacher%
}{%
Cole%
\ \BBA {} Preacher%
}{%
{\protect \APACyear {2014}}%
}]{%
cole&preacher.2014}
\APACinsertmetastar {%
cole&preacher.2014}%
\begin{APACrefauthors}%
Cole, D\BPBI A.%
\BCBT {}\ \BBA {} Preacher, K\BPBI J.%
\end{APACrefauthors}%
\unskip\
\newblock
\APACrefYearMonthDay{2014}{}{}.
\newblock
{\BBOQ}\APACrefatitle {Manifest variable path analysis: Potentially serious and
  misleading consequences due to uncorrected measurement error.} {Manifest
  variable path analysis: Potentially serious and misleading consequences due
  to uncorrected measurement error.}{\BBCQ}
\newblock
\APACjournalVolNumPages{Psychological Methods}{19}{}{300--315}.
\newblock
\begin{APACrefDOI} \doi{10.1037/a0033805} \end{APACrefDOI}
\PrintBackRefs{\CurrentBib}

\bibitem [\protect \citeauthoryear {%
Cronbach%
}{%
Cronbach%
}{%
{\protect \APACyear {1951}}%
}]{%
cronbach.1951}
\APACinsertmetastar {%
cronbach.1951}%
\begin{APACrefauthors}%
Cronbach, L\BPBI J.%
\end{APACrefauthors}%
\unskip\
\newblock
\APACrefYearMonthDay{1951}{}{}.
\newblock
{\BBOQ}\APACrefatitle {Coefficient alpha and the internal structure of tests}
  {Coefficient alpha and the internal structure of tests}.{\BBCQ}
\newblock
\APACjournalVolNumPages{Psychometrika}{16}{3}{297--334}.
\newblock
\begin{APACrefDOI} \doi{10.1007/bf02310555} \end{APACrefDOI}
\PrintBackRefs{\CurrentBib}

\bibitem [\protect \citeauthoryear {%
Cronbach%
\ \BBA {} Meehl%
}{%
Cronbach%
\ \BBA {} Meehl%
}{%
{\protect \APACyear {1955}}%
}]{%
cronbach&meehl.1955}
\APACinsertmetastar {%
cronbach&meehl.1955}%
\begin{APACrefauthors}%
Cronbach, L\BPBI J.%
\BCBT {}\ \BBA {} Meehl, P\BPBI E.%
\end{APACrefauthors}%
\unskip\
\newblock
\APACrefYearMonthDay{1955}{}{}.
\newblock
{\BBOQ}\APACrefatitle {Construct validity in psychological tests.} {Construct
  validity in psychological tests.}{\BBCQ}
\newblock
\APACjournalVolNumPages{Psychological Bulletin}{52}{4}{281--302}.
\newblock
\begin{APACrefDOI} \doi{10.1037/h0040957} \end{APACrefDOI}
\PrintBackRefs{\CurrentBib}

\bibitem [\protect \citeauthoryear {%
Fox%
}{%
Fox%
}{%
{\protect \APACyear {2015}}%
}]{%
fox.2015}
\APACinsertmetastar {%
fox.2015}%
\begin{APACrefauthors}%
Fox, J.%
\end{APACrefauthors}%
\unskip\
\newblock
\APACrefYear{2015}.
\newblock
\APACrefbtitle {Applied Regression Analysis and Generalized Linear Models}
  {Applied regression analysis and generalized linear models}.
\newblock
\APACaddressPublisher{}{SAGE Publications}.
\PrintBackRefs{\CurrentBib}

\bibitem [\protect \citeauthoryear {%
Haberman%
\ \BBA {} Sinharay%
}{%
Haberman%
\ \BBA {} Sinharay%
}{%
{\protect \APACyear {2010}}%
}]{%
haberman&sinharay.2010}
\APACinsertmetastar {%
haberman&sinharay.2010}%
\begin{APACrefauthors}%
Haberman, S\BPBI J.%
\BCBT {}\ \BBA {} Sinharay, S.%
\end{APACrefauthors}%
\unskip\
\newblock
\APACrefYearMonthDay{2010}{}{}.
\newblock
{\BBOQ}\APACrefatitle {Reporting of subscores using multidimensional item
  response theory} {Reporting of subscores using multidimensional item response
  theory}.{\BBCQ}
\newblock
\APACjournalVolNumPages{Psychometrika}{75}{}{209--227}.
\newblock
\begin{APACrefDOI} \doi{10.1007/s11336-010-9158-4} \end{APACrefDOI}
\PrintBackRefs{\CurrentBib}

\bibitem [\protect \citeauthoryear {%
Holland%
}{%
Holland%
}{%
{\protect \APACyear {1990}}%
}]{%
holland.1990}
\APACinsertmetastar {%
holland.1990}%
\begin{APACrefauthors}%
Holland, P\BPBI W.%
\end{APACrefauthors}%
\unskip\
\newblock
\APACrefYearMonthDay{1990}{}{}.
\newblock
{\BBOQ}\APACrefatitle {On the sampling theory roundations of item response
  theory models} {On the sampling theory roundations of item response theory
  models}.{\BBCQ}
\newblock
\APACjournalVolNumPages{Psychometrika}{55}{}{577--601}.
\newblock
\begin{APACrefDOI} \doi{10.1007/bf02294609} \end{APACrefDOI}
\PrintBackRefs{\CurrentBib}

\bibitem [\protect \citeauthoryear {%
Jeon%
\ \BBA {} De~Boeck%
}{%
Jeon%
\ \BBA {} De~Boeck%
}{%
{\protect \APACyear {2016}}%
}]{%
jeon&deboeck.2016}
\APACinsertmetastar {%
jeon&deboeck.2016}%
\begin{APACrefauthors}%
Jeon, M.%
\BCBT {}\ \BBA {} De~Boeck, P.%
\end{APACrefauthors}%
\unskip\
\newblock
\APACrefYearMonthDay{2016}{}{}.
\newblock
{\BBOQ}\APACrefatitle {A generalized item response tree model for psychological
  assessments} {A generalized item response tree model for psychological
  assessments}.{\BBCQ}
\newblock
\APACjournalVolNumPages{Behavior Research Methods}{48}{}{1070--1085}.
\newblock
\begin{APACrefDOI} \doi{10.3758/s13428-015-0631-y} \end{APACrefDOI}
\PrintBackRefs{\CurrentBib}

\bibitem [\protect \citeauthoryear {%
Kim%
}{%
Kim%
}{%
{\protect \APACyear {2012}}%
}]{%
kim.2012}
\APACinsertmetastar {%
kim.2012}%
\begin{APACrefauthors}%
Kim, S.%
\end{APACrefauthors}%
\unskip\
\newblock
\APACrefYearMonthDay{2012}{}{}.
\newblock
{\BBOQ}\APACrefatitle {A note on the reliability coefficients for item response
  model-based ability estimates} {A note on the reliability coefficients for
  item response model-based ability estimates}.{\BBCQ}
\newblock
\APACjournalVolNumPages{Psychometrika}{77}{1}{153--162}.
\newblock
\begin{APACrefDOI} \doi{/10.1007/s11336-011-9238-0} \end{APACrefDOI}
\PrintBackRefs{\CurrentBib}

\bibitem [\protect \citeauthoryear {%
Liu%
\ \BBA {} Pek%
}{%
Liu%
\ \BBA {} Pek%
}{%
{\protect \APACyear {2024}}%
}]{%
liu&pek.2024}
\APACinsertmetastar {%
liu&pek.2024}%
\begin{APACrefauthors}%
Liu, Y.%
\BCBT {}\ \BBA {} Pek, J.%
\end{APACrefauthors}%
\unskip\
\newblock
\APACrefYearMonthDay{2024}{}{}.
\newblock
{\BBOQ}\APACrefatitle {Summed versus estimated factor scores: Considering
  uncertainties when using observed scores} {Summed versus estimated factor
  scores: Considering uncertainties when using observed scores}.{\BBCQ}
\newblock
\APACjournalVolNumPages{Psychological Methods \textnormal{[Advance online
  publication.]}}{}{}{}.
\newblock
\begin{APACrefDOI} \doi{10.1037/met0000644} \end{APACrefDOI}
\PrintBackRefs{\CurrentBib}

\bibitem [\protect \citeauthoryear {%
Liu%
, Pek%
\BCBL {}\ \BBA {} Maydeu-Olivares%
}{%
Liu%
\ \protect \BOthers {.}}{%
{\protect \APACyear {2025}}%
}]{%
Liu2025}
\APACinsertmetastar {%
Liu2025}%
\begin{APACrefauthors}%
Liu, Y.%
, Pek, J.%
\BCBL {}\ \BBA {} Maydeu-Olivares, A.%
\end{APACrefauthors}%
\unskip\
\newblock
\APACrefYearMonthDay{2025}{}{}.
\newblock
{\BBOQ}\APACrefatitle {On a general theoretical framework of reliability} {On a
  general theoretical framework of reliability}.{\BBCQ}
\newblock
\APACjournalVolNumPages{}{78}{}{286--302}.
\newblock
\begin{APACrefDOI} \doi{10.1111/bmsp.12360} \end{APACrefDOI}
\PrintBackRefs{\CurrentBib}

\bibitem [\protect \citeauthoryear {%
Lord%
\ \BBA {} Novick%
}{%
Lord%
\ \BBA {} Novick%
}{%
{\protect \APACyear {1968}}%
}]{%
lord&novick.1968}
\APACinsertmetastar {%
lord&novick.1968}%
\begin{APACrefauthors}%
Lord, F\BPBI M.%
\BCBT {}\ \BBA {} Novick, M\BPBI R.%
\end{APACrefauthors}%
\unskip\
\newblock
\APACrefYear{1968}.
\newblock
\APACrefbtitle {Statistical theories of mental test scores} {Statistical
  theories of mental test scores}.
\newblock
\APACaddressPublisher{}{Addison-Wesley}.
\PrintBackRefs{\CurrentBib}

\bibitem [\protect \citeauthoryear {%
Magnus%
\ \BBA {} Liu%
}{%
Magnus%
\ \BBA {} Liu%
}{%
{\protect \APACyear {2022}}%
}]{%
magnus&liu.2022}
\APACinsertmetastar {%
magnus&liu.2022}%
\begin{APACrefauthors}%
Magnus, B\BPBI E.%
\BCBT {}\ \BBA {} Liu, Y.%
\end{APACrefauthors}%
\unskip\
\newblock
\APACrefYearMonthDay{2022}{}{}.
\newblock
{\BBOQ}\APACrefatitle {Symptom presence and symptom severity as unique
  indicators of psychopathology: An application of multidimensional
  zero-inflated and hurdle graded response models} {Symptom presence and
  symptom severity as unique indicators of psychopathology: An application of
  multidimensional zero-inflated and hurdle graded response models}.{\BBCQ}
\newblock
\APACjournalVolNumPages{Educational and Psychological
  Measurement}{82}{5}{938--966}.
\newblock
\begin{APACrefDOI} \doi{10.1177/0013164421106182} \end{APACrefDOI}
\PrintBackRefs{\CurrentBib}

\bibitem [\protect \citeauthoryear {%
McDonald%
}{%
McDonald%
}{%
{\protect \APACyear {1994}}%
}]{%
mcdonald.1994}
\APACinsertmetastar {%
mcdonald.1994}%
\begin{APACrefauthors}%
McDonald, R\BPBI P.%
\end{APACrefauthors}%
\unskip\
\newblock
\APACrefYearMonthDay{1994}{}{}.
\newblock
{\BBOQ}\APACrefatitle {Testing for approximate dimensionality} {Testing for
  approximate dimensionality}.{\BBCQ}
\newblock
\BIn{} D.~Levault, B\BPBI D.~Zumbo, M\BPBI E.~Gessaroli\BCBL {}\ \BBA {} M\BPBI
  W.~Boss\ (\BEDS), \APACrefbtitle {Modern theories of measurement: Problems
  and issues} {Modern theories of measurement: Problems and issues}\ (\BPGS\
  63--86).
\newblock
\APACaddressPublisher{Ottawa, Canada}{University of Ottawa}.
\PrintBackRefs{\CurrentBib}

\bibitem [\protect \citeauthoryear {%
McDonald%
}{%
McDonald%
}{%
{\protect \APACyear {1999}}%
}]{%
mcdonald.1999}
\APACinsertmetastar {%
mcdonald.1999}%
\begin{APACrefauthors}%
McDonald, R\BPBI P.%
\end{APACrefauthors}%
\unskip\
\newblock
\APACrefYear{1999}.
\newblock
\APACrefbtitle {Test theory: A unified approach} {Test theory: A unified
  approach}.
\newblock
\APACaddressPublisher{Mahwah, NJ}{Lawrence Erlbaum}.
\PrintBackRefs{\CurrentBib}

\bibitem [\protect \citeauthoryear {%
McDonald%
}{%
McDonald%
}{%
{\protect \APACyear {2011}}%
}]{%
mcdonald.2011}
\APACinsertmetastar {%
mcdonald.2011}%
\begin{APACrefauthors}%
McDonald, R\BPBI P.%
\end{APACrefauthors}%
\unskip\
\newblock
\APACrefYearMonthDay{2011}{}{}.
\newblock
{\BBOQ}\APACrefatitle {Measuring latent quantities} {Measuring latent
  quantities}.{\BBCQ}
\newblock
\APACjournalVolNumPages{Psychometrika}{76}{4}{511--536}.
\newblock
\begin{APACrefDOI} \doi{10.1007/s11336-011-9223-7} \end{APACrefDOI}
\PrintBackRefs{\CurrentBib}

\bibitem [\protect \citeauthoryear {%
Revelle%
\ \BBA {} Condon%
}{%
Revelle%
\ \BBA {} Condon%
}{%
{\protect \APACyear {2019}}%
}]{%
revelle&condon.2019}
\APACinsertmetastar {%
revelle&condon.2019}%
\begin{APACrefauthors}%
Revelle, W.%
\BCBT {}\ \BBA {} Condon, D\BPBI M.%
\end{APACrefauthors}%
\unskip\
\newblock
\APACrefYearMonthDay{2019}{}{}.
\newblock
{\BBOQ}\APACrefatitle {Reliability from $\alpha$ to $\omega$: A tutorial.}
  {Reliability from $\alpha$ to $\omega$: A tutorial.}{\BBCQ}
\newblock
\APACjournalVolNumPages{Psychological Assessment}{31}{12}{1395--1411}.
\newblock
\begin{APACrefDOI} \doi{10.1037/pas0000754} \end{APACrefDOI}
\PrintBackRefs{\CurrentBib}

\bibitem [\protect \citeauthoryear {%
Rhemtulla%
, Brosseau-Liard%
\BCBL {}\ \BBA {} Savalei%
}{%
Rhemtulla%
\ \protect \BOthers {.}}{%
{\protect \APACyear {2012}}%
}]{%
rhemtulla.el.al.2012}
\APACinsertmetastar {%
rhemtulla.el.al.2012}%
\begin{APACrefauthors}%
Rhemtulla, M.%
, Brosseau-Liard, P\BPBI {\'E}.%
\BCBL {}\ \BBA {} Savalei, V.%
\end{APACrefauthors}%
\unskip\
\newblock
\APACrefYearMonthDay{2012}{}{}.
\newblock
{\BBOQ}\APACrefatitle {When can categorical variables be treated as continuous?
  A comparison of robust continuous and categorical SEM estimation methods
  under suboptimal conditions.} {When can categorical variables be treated as
  continuous? a comparison of robust continuous and categorical sem estimation
  methods under suboptimal conditions.}{\BBCQ}
\newblock
\APACjournalVolNumPages{Psychological methods}{17}{3}{354}.
\PrintBackRefs{\CurrentBib}

\bibitem [\protect \citeauthoryear {%
Sampson%
}{%
Sampson%
}{%
{\protect \APACyear {1974}}%
}]{%
sampson.1974}
\APACinsertmetastar {%
sampson.1974}%
\begin{APACrefauthors}%
Sampson, A\BPBI R.%
\end{APACrefauthors}%
\unskip\
\newblock
\APACrefYearMonthDay{1974}{}{}.
\newblock
{\BBOQ}\APACrefatitle {A tale of two regressions} {A tale of two
  regressions}.{\BBCQ}
\newblock
\APACjournalVolNumPages{Journal of the American Statistical
  Association}{69}{347}{682--689}.
\newblock
\begin{APACrefDOI} \doi{10.2307/2286002} \end{APACrefDOI}
\PrintBackRefs{\CurrentBib}

\bibitem [\protect \citeauthoryear {%
Sijtsma%
, Ellis%
\BCBL {}\ \BBA {} Borsboom%
}{%
Sijtsma%
\ \protect \BOthers {.}}{%
{\protect \APACyear {2024}}%
}]{%
sijtsma.et.al.2024}
\APACinsertmetastar {%
sijtsma.et.al.2024}%
\begin{APACrefauthors}%
Sijtsma, K.%
, Ellis, J\BPBI L.%
\BCBL {}\ \BBA {} Borsboom, D.%
\end{APACrefauthors}%
\unskip\
\newblock
\APACrefYearMonthDay{2024}{}{}.
\newblock
{\BBOQ}\APACrefatitle {Recognize the Value of the Sum Score, Psychometrics'
  Greatest Accomplishment} {Recognize the value of the sum score,
  psychometrics' greatest accomplishment}.{\BBCQ}
\newblock
\APACjournalVolNumPages{Psychometrika}{89}{1}{84--117}.
\PrintBackRefs{\CurrentBib}

\bibitem [\protect \citeauthoryear {%
Sijtsma%
\ \BBA {} Pfadt%
}{%
Sijtsma%
\ \BBA {} Pfadt%
}{%
{\protect \APACyear {2021}}%
}]{%
sijtsma&pfadt.2021}
\APACinsertmetastar {%
sijtsma&pfadt.2021}%
\begin{APACrefauthors}%
Sijtsma, K.%
\BCBT {}\ \BBA {} Pfadt, J\BPBI M.%
\end{APACrefauthors}%
\unskip\
\newblock
\APACrefYearMonthDay{2021}{}{}.
\newblock
{\BBOQ}\APACrefatitle {Part II: On the use, the misuse, and the very limited
  usefulness of Cronbach’s alpha: Discussing lower bounds and correlated
  errors} {Part ii: On the use, the misuse, and the very limited usefulness of
  cronbach’s alpha: Discussing lower bounds and correlated errors}.{\BBCQ}
\newblock
\APACjournalVolNumPages{Psychometrika}{86}{4}{843--860}.
\newblock
\begin{APACrefDOI} \doi{10.1007/s11336-021-09789-8} \end{APACrefDOI}
\PrintBackRefs{\CurrentBib}

\bibitem [\protect \citeauthoryear {%
Stout%
}{%
Stout%
}{%
{\protect \APACyear {2002}}%
}]{%
stout.2002}
\APACinsertmetastar {%
stout.2002}%
\begin{APACrefauthors}%
Stout, W.%
\end{APACrefauthors}%
\unskip\
\newblock
\APACrefYearMonthDay{2002}{}{}.
\newblock
{\BBOQ}\APACrefatitle {Psychometrics: From practice to theory and back: 15
  years of nonparametric multidimensional {IRT}, {DIF}/test equity, and skills
  diagnostic assessment} {Psychometrics: From practice to theory and back: 15
  years of nonparametric multidimensional {IRT}, {DIF}/test equity, and skills
  diagnostic assessment}.{\BBCQ}
\newblock
\APACjournalVolNumPages{Psychometrika}{67}{}{485--518}.
\newblock
\begin{APACrefDOI} \doi{10.1007/bf02295128} \end{APACrefDOI}
\PrintBackRefs{\CurrentBib}

\bibitem [\protect \citeauthoryear {%
Thissen%
\ \BBA {} Thissen-Roe%
}{%
Thissen%
\ \BBA {} Thissen-Roe%
}{%
{\protect \APACyear {2022}}%
}]{%
thissen&thissen-roe.2022}
\APACinsertmetastar {%
thissen&thissen-roe.2022}%
\begin{APACrefauthors}%
Thissen, D.%
\BCBT {}\ \BBA {} Thissen-Roe, A.%
\end{APACrefauthors}%
\unskip\
\newblock
\APACrefYearMonthDay{2022}{}{}.
\newblock
{\BBOQ}\APACrefatitle {Latent Variable Estimation in Factor Analysis and Item
  Response Theory} {Latent variable estimation in factor analysis and item
  response theory}.{\BBCQ}
\newblock
\APACjournalVolNumPages{Chinese/English Journal of Educational Measurement and
  Evaluation}{3}{3}{Article 1}.
\newblock
\begin{APACrefDOI} \doi{10.59863/optz4045} \end{APACrefDOI}
\PrintBackRefs{\CurrentBib}

\bibitem [\protect \citeauthoryear {%
Thissen%
\ \BBA {} Wainer%
}{%
Thissen%
\ \BBA {} Wainer%
}{%
{\protect \APACyear {2001}}%
}]{%
thissen&wainer.2001}
\APACinsertmetastar {%
thissen&wainer.2001}%
\begin{APACrefauthors}%
Thissen, D.%
\BCBT {}\ \BBA {} Wainer, H.%
\end{APACrefauthors}%
\unskip\
\newblock
\APACrefYear{2001}.
\newblock
\APACrefbtitle {Test scoring} {Test scoring}.
\newblock
\APACaddressPublisher{Mahwah, NJ}{Lawrence Erlbaum}.
\PrintBackRefs{\CurrentBib}

\bibitem [\protect \citeauthoryear {%
Thomson%
}{%
Thomson%
}{%
{\protect \APACyear {1936}}%
}]{%
thomson.1936}
\APACinsertmetastar {%
thomson.1936}%
\begin{APACrefauthors}%
Thomson, G\BPBI H.%
\end{APACrefauthors}%
\unskip\
\newblock
\APACrefYearMonthDay{1936}{}{}.
\newblock
{\BBOQ}\APACrefatitle {Some points of mathematical technique in the factorial
  analysis of ability.} {Some points of mathematical technique in the factorial
  analysis of ability.}{\BBCQ}
\newblock
\APACjournalVolNumPages{}{}{27}{36--54}.
\newblock
\begin{APACrefDOI} \doi{10.1037/h0062007} \end{APACrefDOI}
\PrintBackRefs{\CurrentBib}

\bibitem [\protect \citeauthoryear {%
Thurstone%
}{%
Thurstone%
}{%
{\protect \APACyear {1935}}%
}]{%
thurstone.1935}
\APACinsertmetastar {%
thurstone.1935}%
\begin{APACrefauthors}%
Thurstone, L\BPBI L.%
\end{APACrefauthors}%
\unskip\
\newblock
\APACrefYear{1935}.
\newblock
\APACrefbtitle {The vectors on mind} {The vectors on mind}.
\newblock
\APACaddressPublisher{}{University of Chicago Press}.
\PrintBackRefs{\CurrentBib}

\bibitem [\protect \citeauthoryear {%
Wood%
}{%
Wood%
}{%
{\protect \APACyear {2003}}%
}]{%
wood.2003}
\APACinsertmetastar {%
wood.2003}%
\begin{APACrefauthors}%
Wood, S\BPBI N.%
\end{APACrefauthors}%
\unskip\
\newblock
\APACrefYearMonthDay{2003}{}{}.
\newblock
{\BBOQ}\APACrefatitle {Thin plate regression splines} {Thin plate regression
  splines}.{\BBCQ}
\newblock
\APACjournalVolNumPages{Journal of the Royal Statistical Society Series B:
  Statistical Methodology}{65}{1}{95--114}.
\newblock
\begin{APACrefDOI} \doi{10.1111/1467-9868.00374} \end{APACrefDOI}
\PrintBackRefs{\CurrentBib}

\end{thebibliography}

\includepdf[pages=-]{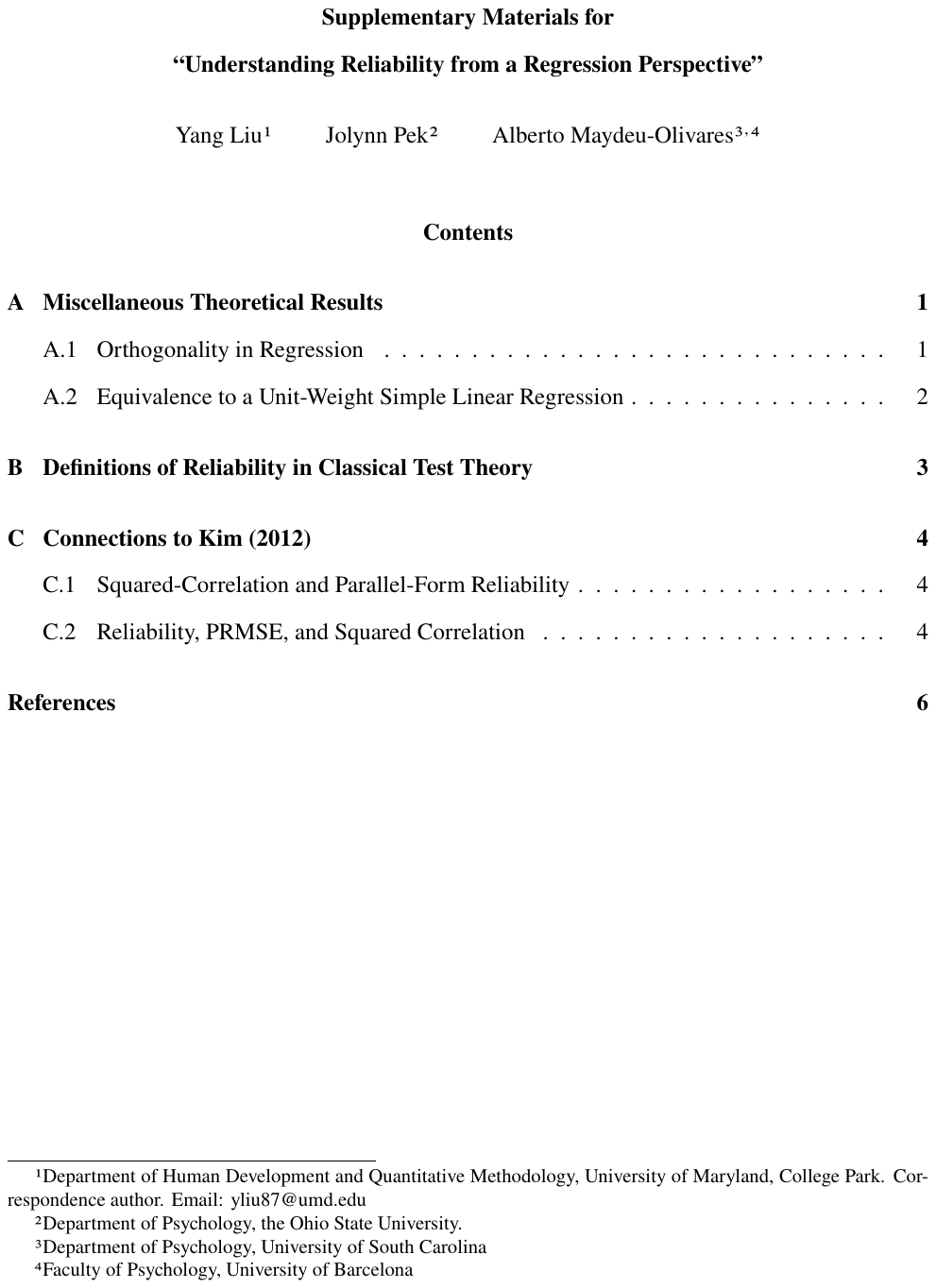}

\end{document}